\documentclass[useAMS,usenatbib]{mn2e}

\usepackage{graphicx}

\newcommand{\kms}{km s$^{-1}$}
\newcommand{\cmN}{cm$^{-2}$}
\newcommand{\cmn}{cm$^{-3}$}
\newcommand{\msun}{M$_{\odot}$}

\newcommand{\lam}{$\lambda$}
\newcommand{\aj}{AJ} 
\newcommand{\mnras}{MNRAS} 
\newcommand{\apj}{ApJ} 
\newcommand{\apjl}{ApJ} 
\newcommand{\apjs}{ApJS} 
\newcommand{\aap}{A\&A} 
\newcommand{\aaps}{A\&AS} 
\newcommand{\araa}{ARA\&A} 

\newcommand{\pasp}{PASP} 

\newcommand{\nat}{Nat} 
\newcommand{\heii}{\mbox{He\,{\sc ii}}}

\newcommand{\civ}{\mbox{C\,{\sc iv}}}
\newcommand{\ciii}{\mbox{C\,{\sc iii}}}
\newcommand{\cii}{\mbox{C\,{\sc ii}}}
\newcommand{\siiv}{\mbox{Si\,{\sc iv}}}
\newcommand{\siiii}{\mbox{Si\,{\sc iii}}}
\newcommand{\siii}{\mbox{Si\,{\sc ii}}}
\newcommand{\nv}{\mbox{N\,{\sc v}}}
\newcommand{\niv}{\mbox{N\,{\sc iv}}}
\newcommand{\niii}{\mbox{N\,{\sc iii}}}
\newcommand{\ovii}{\mbox{O\,{\sc vii}}}
\newcommand{\ovi}{\mbox{O\,{\sc vi}}}
\newcommand{\ov}{\mbox{O\,{\sc v}}}
\newcommand{\oiv}{\mbox{O\,{\sc iv}}}
\newcommand{\oiii}{\mbox{O\,{\sc iii}}}

\newcommand{\oi}{\mbox{O\,{\sc i}}}
\newcommand{\lya}{\mbox{Ly$\alpha$}}
\newcommand{\lyb}{\mbox{Ly$\beta$}}
\newcommand{\hi}{\mbox{H\,{\sc i}}}

\title[A High-Velocity Narrow Absorption Line Outflow]{A High-Velocity    
Narrow Absorption Line Outflow in the Quasar J212329.46-005052.9}
\author[F. Hamann et al.]{F. Hamann$^{1}$\thanks{E-mail:
hamann@astro.ufl.edu (FH)}, N. Kanekar$^{2}$, J. X. Prochaska$^{3,4}$, 
M. T. Murphy$^{5}$, S. Ellison$^{6}$, \newauthor A. L. Malec$^{5}$, 
N. Milutinovic$^{6}$, and W. Ubachs$^7$\\
$^{1}$Department of Astronomy, University of Florida, Gainesville, FL 
32611-2055, USA\\
$^{2}$Ramanujan Fellow, 
National Centre for Radio Astrophysics, Tata Institute of 
Fundamental Research, Ganeshkhind, Pune 411007, India\\
$^{3}$Department of Astronomy and Astrophysics,
University of California, Santa Cruz, Santa Cruz, CA 95064, USA\\
$^{4}$University of California Observatories -
Lick Observatory, University of California, Santa Cruz, CA 95064, USA\\
$^{5}$Centre for Astrophysics \& Supercomputing,
Swinburne University of Technology, Melbourne, 
Victoria 3122, Australia\\
$^{6}$Department of Physics \& Astronomy, University of Victoria, 
Victoria, BC, V8P 1A1, Canada\\
$^{7}$Laser Centre, VU University, De Boelelaan 1081, 1081 HV Amsterdam, The Netherlands}

\begin{document}

\date{Accepted xxx. Received xxx}

\pagerange{\pageref{firstpage}--\pageref{lastpage}} \pubyear{2008}

\maketitle

\label{firstpage}

\begin{abstract}

We report on the discovery of a high-velocity narrow absorption 
line outflow in the redshift 2.3 quasar J212329.46-005052.9. 
Five distinct outflow systems are detected with velocity 
shifts from $-$9710 to $-$14,050 \kms\ and \civ\ \lam\lam 1548,1551 
line widths of FWHM $\approx$ 62 to 164 \kms . 
This outflow is remarkable for having high speeds and a degree of  
ionization similar to broad absorption line (BAL) flows, but line 
widths roughly 100 times narrower than BALs and no apparent X-ray 
absorption. This is also, to our knowledge, the highest-velocity 
narrow absorption line system confirmed to be in a quasar outflow 
by all three indicators of line variability, smooth super-thermal 
line profiles and doublet ratios that require partial covering of 
the quasar continuum source. All five systems have stronger absorption 
in \ovi\ \lam\lam 1032,1038 than \civ\ with no lower ionization 
metal lines detected. Their line variabilities also appear 
coordinated, with each system showing larger changes in \civ\ 
than \ovi\ and line strength variations accompanied by nearly 
commensurate changes in the absorber covering fractions. 
The metallicity is approximately twice solar. 

These data require five distinct outflow structures with 
similar kinematics, physical conditions and characteristic 
sizes of order 0.01-0.02 pc (based on partial covering). 
The coordinated line variations, occurring on time scales 
$\leq$0.63 yr (quasar frame), are best 
explained by global changes in the outflow ionization 
caused by changes in the quasar's ionizing flux. 
An upper limit on the acceleration, $\la$3 \kms\ yr$^{-1}$, 
is consistent with blobs of gas that are 
gravitationally unbound and coasting freely $\ga$5 pc from the 
central black hole. Additional constraints from the variability 
time indicate that the full range of plausible distances 
is $5\la R\la 1100$ pc. However, if these small absorbing structures 
were created in the inner flow, they should be near the $\sim$5 pc 
minimum radius because they can travel just 
a few pc before dissipating (without external confinement). 
An apparent double line-lock in \civ\ suggests that the 
flow was radiatively accelerated and its present trajectory 
is within $\sim$16$^o$ of the radial (line-of-sight) 
direction. The absence of strong X-ray absorption shows that 
radiative shielding in the far-UV and X-rays is not needed 
to maintain moderate BAL-like ionizations and 
therefore, apparently, it is not needed to facilitate the 
radiative acceleration to high speeds. 
We argue that the ionization is moderated, instead, by 
high gas densities in small outflow sub-structures. 
Finally, we estimate that the kinetic energy yield from this 
outflow is at least two orders of magnitude too low to be important 
for feedback to the host galaxy's evolution. 
\end{abstract}


\section{Introduction}

High-velocity outflows from quasars appear to be a natural part of 
the accretion process. They are detected most conspicuously 
via broad absorption lines (BALs) in quasar spectra, with velocity 
widths $>$2000 \kms\ \citep[by definition,][]{Weymann91} and  
flow speeds from a few thousand to roughly 10,000-30,000 \kms\ 
\citep[see also][]{Korista93,Trump06}. However, quasar outflows  
can also produce narrow absorption lines (NALs), with velocity widths 
less than a few hundred \kms , as well as mini-BALs, which are 
loosely defined as intermediate between NALs and BALs 
\citep[e.g.,][]{Hamann04,Nestor08,Paola10}. 
These narrower outflow lines are more common than BALs 
\citep{Paola10,Gibson09b,Nestor08,Wild08,Misawa07a,Richards01a}. 
Overall it is estimated that 
$>$50\% of optically-selected quasars have at least one 
type of outflow absorption line in their 
rest-frame UV spectra \citep{Ganguly08,Paola10}. 
The flows themselves are probably present in all quasars if, as 
expected, the outflowing gas covers just a fraction of the sky 
as seen from the central emission source \citep[e.g.,][]{Hamann93}.  

Quasar outflows have gained attention recently as a mechanism that 
can physically couple quasars to the evolution of their host 
galaxies. In particular, some models of galaxy evolution invoke the 
kinetic energy ``feedback" from an accreting super-massive 
black hole (SMBH) to regulate both the star formation in the host 
galaxies and the infall of matter toward the central SMBH 
\citep[][and refs. therein]{Kauffmann00,Granato04,DiMatteo05,Hopkins10}. 
This type of coupling via feedback could provide a natural explanation 
for the observed mass correlation between SMBHs and their host 
galaxy spheroids \citep{Tremaine02,Marconi03,Haring04}. 
The kinetic energy 
luminosity, $L_K$, needed for significant feedback from a quasar 
outflow is believed to be just a few percent of the quasar bolometric 
(photon) luminosity, e.g., $L_K/L\sim 5\%$ 
\citep{Scannapieco04,DiMatteo05,Prochaska09}. 

Unfortunately, many aspects of the energetics 
and physical properties of quasar outflows remain 
poorly understood. Computational models 
aimed mainly at BALs attribute the flows to gas 
lifted off of the accretion disk and driven outward 
to high speeds by radiation pressure or magneto-centrifugal forces 
\citep{Murray95,Murray97,Proga04,Everett05}. Radiative forces 
are expected to dominate in quasars, where the bolometric 
luminosities are a substantial fraction of the 
Eddington luminosity \citep{Everett05}. 

One essential requirement for radiative driving is 
that the outflow is not too highly ionized, so it can 
maintain significant line and bound-free continuum opacities 
across the main acceleration region. This is not trivial for 
flows launched from the inner accretion 
disk, near the quasar's intense source of ionizing radiation. 
\cite{Murray95} proposed that the outflow ionization is 
moderated by a high column density  
of roughly stationary material that serves as a radiative 
shield at the base of the flow \cite[also][]{Murray97}. 
The shielding medium is itself 
too ionized and too transparent to be driven radiatively to 
high speeds, but it might have enough continuous opacity in 
the far-UV and X-rays to lower the ionization and 
thereby facilitate the acceleration of the gas 
directly behind it (i.e., farther from the emission 
source). This scenario is supported by observations showing that 
BAL quasars are heavily obscured in X-rays, 
consistent with strong radiative shielding  
\citep{Green96,Mathur00,Gallagher02,Gallagher06}. 
However, other observations indicate that quasars with 
NAL or mini-BAL outflows have substantially less  
X-ray absorption \citep{Chartas09,Gibson09a,Misawa08}. This 
disparity in the X-ray results, between BALs on the one hand 
and NALs and mini-BALs on the other, presents  
an important challenge to the paradigm of radiative 
acceleration behind an X-ray/far-UV shield (see also \S4.5  
below). 

A central problem is our poor understanding of the 
physical relationships between BAL, NAL and mini-BAL outflows. 
One possibility is that the different line types are simply 
different manifestations of a single outflow phenomenon viewed at 
different angles. For example, it is often supposed that BALs  
form in the main body of the outflow near the accretion disk 
plane, while NALs and mini-BALs form along 
sightlines that skim the edges of the BAL flow at higher 
latitudes above the disk \citep{Ganguly01,Chartas09}. 
There could also be evolutionary effects. For example,  
NALs and mini-BALs might represent the tentative beginning or 
end stages of a more powerful BAL outflow phase \citep{Hamann08}. 
There is some evidence for outflow evolution 
in that a particular variety of low-ionization BALs (the so-called 
FeLoBALs) is found preferentially in dusty young host galaxies 
with high star formation rates \citep{Farrah07}. Variability studies 
have shown further that BALs and mini-BALs in quasar spectra can appear, 
disappear or swap identities (where a BAL becomes a mini-BAL or vice 
versa) on time scales of months to years in the quasar rest 
frame \citep{Hamann08,Gibson08,Leighly09,Capellupo10,Paola10,Gibson10}. 
The true relationship between outflow NALs, BALs and mini-BALs probably 
involves an amalgam of orientation and time-dependent effects. 

To build a more complete picture of quasar outflows, we need 
better observational constraints on the physical properties of each 
outflow type. NAL outflows present a unique challenge because 
they are difficult to identify. Most 
narrow absorption lines in quasar spectra form in cosmologically 
intervening gas that has no relationship to the 
background quasar. Statistical studies that examine the 
velocity distributions of the narrow metal lines 
\citep{Nestor08,Wild08}, or correlations between the 
incidence of these lines and the properties of the background 
quasars \citep{Richards01a}, indicate that significant fractions  
of the narrow metal-line systems do arise in quasar outflows. 
However, to identify individual NAL outflow systems 
we need more information. The most commonly used indicators of an 
outflow origin are line variability, resolved absorption profiles 
that are significantly broad and smooth compared to thermal 
line widths, or line strength ratios in multiplets that reveal 
partial line-of-sight covering of the background 
light source \citep{Hamann97,Barlow97,Misawa07c,Hamann10,Simon10}. 

Here we discuss a complex of five outflow NAL systems at velocities 
from $-$9710 to $-$14,050 \kms\ in the redshift $\sim$2.3 quasar 
J212329.46-005052.9 (hereafter J2123-0050). To our knowledge, 
these are the highest velocity NALs whose  
outflow nature is confirmed by all three indicators of 
variability, partial covering and smooth super-thermal line 
profiles. We selected this quasar from the Sloan Digital Sky 
Survey (SDSS) as part of a larger study of intervening, metal-strong 
damped Ly$\alpha$ (DLA) and sub-DLA absorption line systems 
\citep{Herbert-Fort06}. Follow-up observations 
\citep{Kaplan10,Malec10,Milutinovic10}
revealed the variability and outflow origin of the NAL systems 
discussed here. In the sections below we 
describe the observations (\S2), present measurements and 
analysis of the outflow lines (\S3), and discuss the interpretation 
and broader implications of our results (\S4). 
We provide a detailed general summary in \S5. 
Throughout this paper, we adopt a cosmology with 
$H_o = 70$ \kms\ Mpc, $\Omega_M = 0.3$ and $\Omega_{\Lambda}=0.7$.  
 
\section[]{Observations \& Other Data}

Table 1 summarizes the observations used in this 
study. Columns 1 and 2 give the fractional years and calendar 
dates of the observations. Columns 3-5 list the telescope, 
spectral resolution and wavelength coverage, 
$\Delta\lambda$, respectively. 
We obtained the 2002.68 spectrum from the SDSS archives fully 
reduced and calibrated. The fluxes provided by the SDSS 
are expected to be accurate to within a few percent 
\citep{Adelman08}. Unfortunately, these are the only 
reliable absolute fluxes available for our spectra. 

We observed J2123-0050 at the MMT observatory using the MMT  
Spectrograph in 2006.71 and 2009.40. The resolutions and 
wavelength coverages listed in Table 1 were obtained using a 
1.5$^{\prime\prime}$ wide entrance slit together with the 
800 groove/mm grating in the blue channel. We extracted and 
calibrated these spectra with standard techniques using the Low Redux 
Pipeline\footnote{http://www.ucolick.org/$\sim$xavier/LowRedux/index.html} 
software package. We performed relative flux calibrations 
using spectrophotometric standards observed on the same night. 

The emission-line redshift of J2123-0050 reported by the SDSS, 
$z_{e}=2.2614$, was modified recently to $2.2686 \pm 0.0003$ 
by \cite{Hewett10}. These redshifts match the 
peak position of the broad \civ\ \lam\lam 1548,1551 
emission line quite well. However, the lower ionization  
lines \oi\ \lam 1303 and \cii\ \lam 1336 appear 
at a slightly higher redshift that is probably 
closer to the true quasar systemic 
\citep[within $\sim$100 \kms ;][]{Tytler92,Shen07}. We therefore 
estimate the redshift from the \oi\ and \cii\ lines as 
follows. First we average together the SDSS and MMT 2006.71 
spectra to maximize the signal-to-noise ratio. Then we manually 
interpolate across the tops of any narrow absorption lines 
to eliminate these features (e.g., in the \oi\ emission line 
profile), and calculate the \oi\ and \cii\ emission line 
centroids. This yields an average redshift for the two lines of  
$z_e=2.278\pm 0.002$ \citep[using laboratory 
wavelengths from][]{Verner96}. Measurements based on the 
line peaks or the centroids of only the upper halves of the profiles 
yield essentially the same result. Also, redshifts determined 
separately for the two lines agree within the measurement uncertainty 
quoted above. We will adopt the redshift $z_e=2.278\pm 0.002$ 
for the quasar systemic throughout 
the remainder of this paper. 

\begin{table}
 \centering
 \begin{minipage}{80mm}
  \caption{Observation Summary}
  \begin{tabular}{@{}llccc@{}}
  \hline
 Year & Date& Telesc. & Resol. & $\Delta\lambda$\\
 & (yyyy/mm/dd)& & (\kms )& (\AA )\\
\hline
2002.68& 2002/09/06& SDSS& 150& 3805-9205\\
2006.64& 2006/08/20& Keck& 2.7& 3057-5896\\
2006.71& 2006/09/16& MMT& 220& 3050-4988\\
2008.61& 2008/08/10& VLT& 5.5& 3047-3870\\
~~~~``& ~~~~~~``& ``& 4.5& 4622-9466\\
2008.65& 2008/08/27& VLT& ``&  ``\\
2008.66& 2008/08/30-31& VLT& ``& ``\\
2008.73& 2008/09/22& VLT& ``& ``\\
2008.73& 2008/09/25& VLT& ``& ``\\
2009.40& 2009/05/25& MMT& 220& 3329--5968\\

\hline
\end{tabular}
\end{minipage}
\end{table}
 
We observed J2123-0050 at high spectral 
resolutions at the W. M. Keck observatory using the High 
Resolution Echelle Spectrograph (HIRES) and at the Very 
Large Telescope (VLT) using the Ultraviolet and Visual Echelle 
Spectrograph (UVES). We extracted and calibrated the Keck spectra 
using the HIRES reduction package 
HIRedux\footnote{http://www.ucolick.org/$\sim$xavier/HIRedux/index.html}. 
More information 
about the Keck/HIRES observations and data reductions can 
be found in \cite{Milutinovic10}. We obtained the VLT/UVES 
spectra in separate blue and red channels covering the wavelengths 
indicated in Table 1. These data were reduced by the UVES 
data reduction pipeline. The final wavelength scales in both the 
Keck and VLT spectra are in the vacuum heliocentric frame. 
We combined the UVES spectra obtained in different echelle orders 
and in different exposures using the customized code  
UVES\_popler\footnote{http://astronomy.swin.edu.au/$\sim$mmurphy/UVES\_popler.html}. The VLT/UVES spectra 
were obtained on several dates during a $\sim$6 week 
period in 2008 (Table 1). Careful examination of these 
spectra shows that there are no significant 
differences in any of the absorption lines within 
this data set. Hereafter, we discuss only the average 
VLT spectrum referred to by the average date, 2008.67. 
This spectrum 
represents a total of 11.3 hrs of total exposure time on 
the quasar.

To measure the absorption lines, we define pseudo-continua in 
all of the spectra by manually drawing 
a smooth curve that matches the data in every important 
detail while extrapolating over the tops of the absorption lines. 
We then divide the original data by these pseudo-continua to obtain  
normalized spectra for our analysis below. 
This procedure works well redward of the the \lya\ emission 
line ($\lambda > 3985$ \AA\ observed) in all of our data. 
However, in the lower resolution SDSS 
and MMT data, the blending of lines in the \lya\ forest is too 
severe to yield useful results. We therefore exclude the 
SDSS and MMT data at these short wavelengths from our analysis.

\subsection{Quasar Properties}

The photometric magnitudes reported by the SDSS  
(e.g., $r=16.4$ and $i=16.3$) indicate that J2123-0050 
is one of the most luminous quasars known. 
\citep[Inspection of the SDSS images reveals no 
evidence for gravitational lensing 
that might artificially boost the observed fluxes,][]{Just07}. 
We estimate the bolometric luminosity of 
J2123-0050 to be $L=8.4\times 10^{47}$ ergs s$^{-1}$ 
based on the specific luminosity at 1450 \AA\ in the rest 
frame of $\lambda L_{\lambda}(1450{\rm \AA }) \approx 1.9 
\times 10^{47}$ ergs s$^{-1}$ (see Figure 2 below)   
and a bolometric correction factor of $L = 4.4\, 
\lambda L_{\lambda}$(1450\AA ) \citep[see Appendix, also][]{Warner04}. 
Combining this luminosity with our own measurement 
of the \civ\ emission line width, FWHM $\approx 8750$ \kms\ 
from the SDSS spectrum, yields an estimate of the 
black hole mass, $M_{BH}\approx 2\times 10^{10}$ M$_{\odot}$ 
\citep{Vestergaard06}, and 
the Eddington ratio, $L/L_{E}\approx 0.4$, where $L_E$ is 
the Eddington luminosity. These values of $M_{BH}$ and 
$L/L_E$ are within the nominal range of other high-luminosity 
quasars \citep{Shemmer04,Netzer07}.

Some of our analysis in \S3 and \S4 below relies on size 
estimates for the quasar emission sources. If  
the UV continuum source is a geometrically thin accretion disk 
that emits like a blackbody at every radius, then the size of 
the continuum emitter at different wavelengths depends only 
on the bolometric luminosity and the radial run 
of temperature, $T(r)$, in the disk \citep{Peterson97}. The 
luminosity of J2123-0050 combined with $T(r)\propto r^{-0.57}$ 
\citep{Gaskell08} indicates continuum 
source diameters of $\sim$0.026 pc at 1550 \AA\ and 
$\sim$0.013 pc at 1034 \AA\ \citep{Hamann10}. The radius 
of the \civ\ broad emission line region (BELR) should be 
roughly $R_{BELR}\sim 0.65$ pc based on the empirical scaling 
relation with luminosity \citep{Bentz07} 
as formulated by \cite{Hamann10}.

\subsection{Other Data}

\cite{Just07} note that J2123-0050 is radio quiet
based on a non-detection in the FIRST radio survey 
\citep{Becker95}. They also report that J2123-0050 has a relatively 
``soft'' X-ray spectrum, with a powerlaw slope of $\alpha < -1.1$ 
($f_{\nu}\propto \nu^{\alpha}$) from roughly 0.5 to 8 keV (observed 
frame) and no evidence for continuous X-ray absorption in the 
quasar environment. The short exposure time of the X-ray 
observations does not provide a strong 
upper limit on the column density of X-ray absorption. 
However, the X-ray characteristics of J2123-0050 are 
like other luminous radio-quiet quasars, for which 
\cite{Just07} derive an upper limit of 
$N_H \la 2\times 10^{21}$ \cmN\ on the X-ray absorbing column 
using coadded spectra of quasars in their sample. 
Consistent with a lack of X-ray absorption, 
J2123-0050 has a 2-point power law index of $\alpha_{ox} = 1.91$ 
between 2500 \AA\ and 2 keV in the quasar frame, which is in the 
nominal range for luminous radio-quiet quasars with no 
X-ray absorption \citep{Just07,Steffen06}.
  
\section[]{Analysis \& Results}

\subsection{Defining the Variable Systems}
 
Figure 1 shows the five variable \civ\ \lam\lam 1548,1551 doublets 
(labeled A-E) in the normalized Keck 2006.64 and VLT 2008.67 
spectra. Significant changes in the line strengths are clearly 
evident. It is also evident that the changes were well 
coordinated between the five systems (see also \S3.2 below). 
 \begin{figure*}
 \includegraphics[scale=0.6,angle=-90.0]{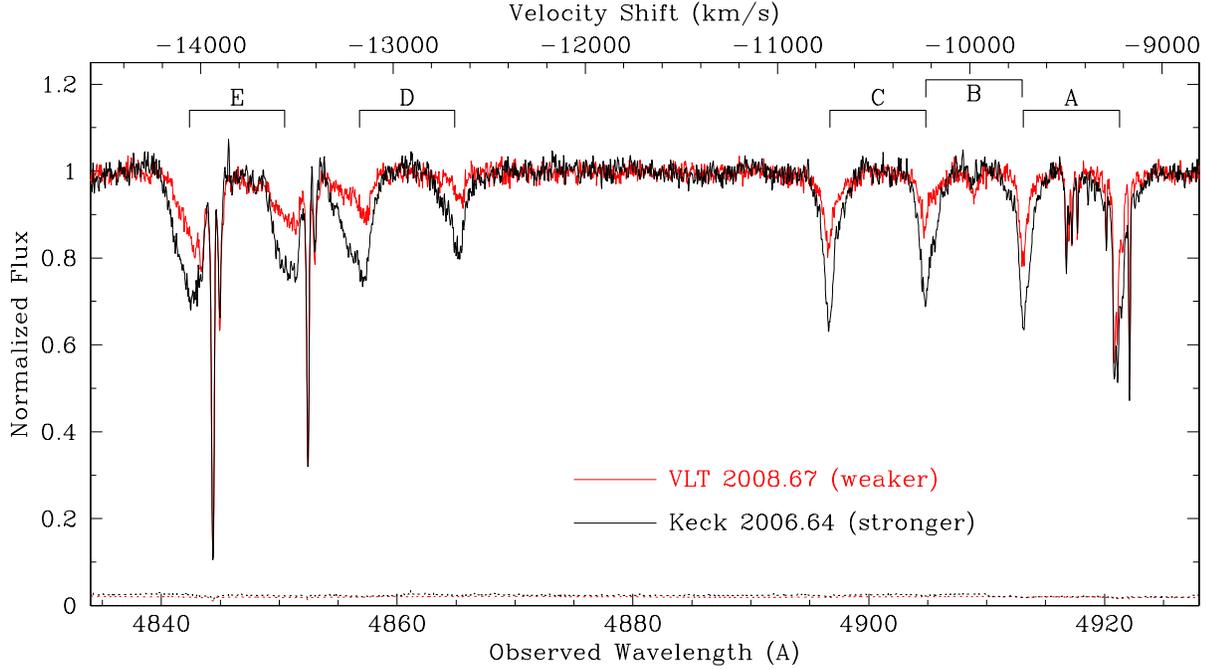}
\vspace{-3pt}
 \caption{Normalized spectra of J2123-0050 showing the variable 
\civ\ systems 
 A-E in the two high-resolution observations: 2006.64 (black curve) 
 and 2008.67 (red). The 
 velocity scale across the top applies to the short 
 wavelength lines in the doublet, \civ\ \lam 1548, relative to the 
 emission redshift, $z_e = 2.278$. The 1$\sigma$ variance spectra 
are plotted as dotted black and red curves across the bottom. 
Much narrower absorption lines not labeled are cosmologically 
intervening and unrelated to the variable outflow systems. 
  }
\end{figure*}

Figure 2 shows the variable \civ\ NALs in relation to other 
features in the J2123-0050 spectrum. The spectrum shown in 
this figure represents the product of the normalized 
Keck 2006.64 data times the pseudo-continuum drawn through 
the SDSS 2002.68 spectrum. A small featureless gap in the Keck 
wavelength coverage around 4970 \AA\ (observed) is filled using 
the VLT spectrum. The result in Figure 2 thus has 
the fluxes, broad emission lines and overall shape matching the SDSS 
data in 2002.68, but with the absorption lines measured at much 
higher resolution at Keck in 2006.64. 
\begin{figure*}
 \includegraphics[scale=0.8]{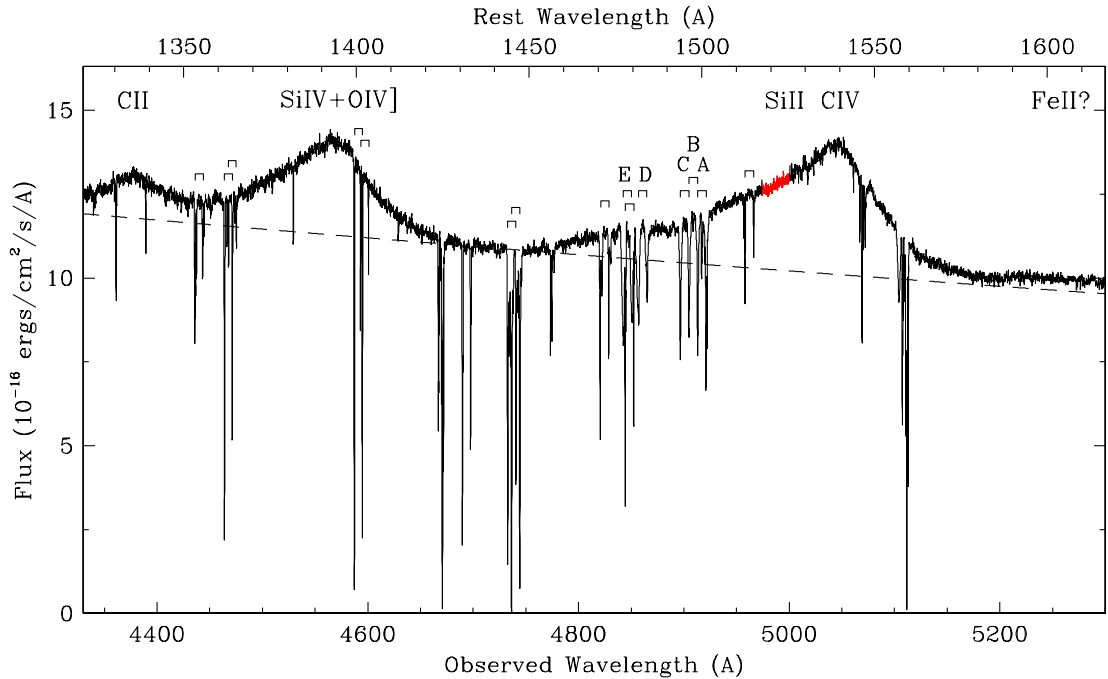}
 \vspace{-8pt}
 \caption{Synthesized spectrum of J2123-0050 showing the absorption 
 lines in the Keck 2006.64 data relative to the 
 broad emission lines and overall spectral shape defined 
 by the SDSS measurement in 2002.68. All of the 
 \civ\ \lam\lam 1548,1551 absorption doublets detected in this 
 wavelength range are marked by open brackets above the spectrum. 
 The variable outflow systems are labeled A-E. 
 Various broad emission lines are labeled across the top. 
 A small featureless segment near 4970 \AA\ 
 observed (drawn in red) uses the average VLT 2008.67 spectrum to 
 fill a small gap in the Keck wavelength coverage. The Keck and VLT 
 spectra are shown after binomial smoothing to improve the 
presentation. The dashed curve is a powerlaw fit to the 
underlying quasar continuum.}
\end{figure*}

There are a number of other narrow \civ\ systems 
scattered across the J2123-0050 spectrum. They include one system at 
a smaller velocity shift than the variable lines 
(at $\sim$4960 \AA\ observed) and another that is blended with the 
variable system E (see also Figure 1). Careful comparisons between 
the Keck and VLT spectra shows that none of these other narrow \civ\ 
systems varied significantly (e.g., by more than a few percent in their 
rest equivalent widths) between 
the 2006.64 and 2008.67 observing epochs. Several \civ\ systems 
measured in the Keck spectrum appear at wavelengths not covered 
by the VLT data. For these systems, we searched for variability 
in all of the other data sets listed in Table 1. We again find 
no evidence for variability in the other lines, in this case above 
an estimated sensitivity threshold of $\sim$20\% in the integrated 
line strengths. The only variable \civ\ lines detected in our data 
are the five systems labeled A-E in Figures 1 and 2. 

One important property of the variable \civ\ lines is that they 
all reach signficantly below the continuum 
shown by the dashed curve in Figure 2. We estimate this 
continuum based on a single powerlaw fit constrained by the 
measured flux 
in narrow wavelength intervals (1437-1448 \AA , 1675-1700 \AA\ 
and 1990-2030 \AA\ rest) that avoid the broad emission lines. 
The powerlaw index of the fit is $\alpha = -0.67$ (for 
$f_{\nu}\propto \nu^{\alpha}$). The fact that the variable 
systems signficantly absorb the quasar continuum is important 
because the partial line-of-sight covering deduced from the line ratios 
(\S3.3 and \S3.4 below) applies to the continuum source and 
not (or not only) to the much larger BELR (\S2.1). 

Finally, we search for lines of other ions in the five variable systems 
using both the Keck 2006.64 and VLT 2008.67 data. This search 
is complicated for wavelengths below $\sim$1270 \AA\ in the absorber  
frame because of contamination by the dense forest of unrelated 
(intervening) \lya\ absorption lines. Nonetheless, we find strong 
absorption in \ovi\ \lam\lam 1032,1038 in all five variable systems, 
probable absorption in \nv\ \lam\lam 1239,1242 in some of those 
systems, and a secure detection of \lya\ in system C. 
We note that the detection of \ovi\ in system B 
confirms the reality of this system, which is not obvious from 
\civ\ alone because of the blending with systems A and C 
(see Fig. 2 and Fig. 4 below). No other lines are detected in 
the variable systems, including specifically \lyb\ and low-ionization 
metal lines such as \cii\ \lam 1336, \ciii\ \lam 977, 
\siiii\ \lam 1206 and \siiv\ \lam\lam 1394,1403. 

\subsection{Line Measurements \& Variability Properties}

Table 2 lists several parameters of the variable \civ\ lines 
A--E measured from the normalized Keck 2006.64 and VLT 2008.67 
spectra. These results were obtained 
using cursor commands in the IRAF\footnote{IRAF is 
distributed by the National Optical Astronomy Observatory, 
which is operated by the Association of Universities 
for Research in Astronomy (AURA) under cooperative 
agreement with the National Science Foundation.}  
software package. For comparison, Table 2 also lists 
measurements (from the Keck spectrum only) for the 
three non-variable \civ\ systems that are closest in velocity 
to systems A-E. Column 1 in the table gives the variable system 
name, where A+B and C+B refer to the blended \civ\ features. 
Column 2 identifies the line 
within the \civ\ doublet. For the blends A+B and C+B, the 
doublet line listed pertains to systems A and C, respectively, 
which appear to dominate the absorption based on our line 
fits in \S3.3 below. Columns 3 and 4 give the observed 
wavelengths ($\lambda_{obs}$, vacuum heliocentric) and 
redshifts ($z_a$) of the line 
centroids. Column 5 lists the centroid velocities ($v$) 
relative to $z_e = 2.278$. Columns 6 and 7 give the rest 
equivalent widths (REW) and the full widths at half minimum (FWHM). 
The non-variable \civ\ systems at $z_a = 2.1291$ and $z_a = 2.1139$ 
each consist of two or more blended sub-components. 
The table lists the total REWs for these blends while the 
FWHM pertains to the strongest single line in the blend. 

\begin{table*}
 \centering
\begin{minipage}{166mm}
  \caption{\civ\ Line Measurements}
  \begin{tabular}{@{}ccccccccccccc@{}}
  \hline
 & &   \multicolumn{5}{c}{--------------------- 2006.64 Keck ---------------------}& & 
 \multicolumn{5}{c}{--------------------- 2008.67 VLT ------------------------}\\
  System & Line& $\lambda_{obs}$& $z_a$& $v$ & REW& FWHM& & 
  $\lambda_{obs}$& $z_a$& $v$& REW& FWHM \\
& & (\AA )& & (\kms )& (\AA )& (\kms )& & (\AA )& & (\kms )& (\AA )& (\kms )\\
\hline
 
A& 1551&  4921.3& 2.17343&  $-$9716& 0.150&  62& & 4921.3& 2.17342&  $-$9717& 0.063&  55\\
A+B&  1548& 4913.2& 2.17349&  $-$9710& 0.156&  66& & 4913.2& 2.17348&  $-$9711& 0.061&  50\\
C+B&  1551& 4905.0& 2.16291&  $-$10710& 0.144&  85& & 4905.0& 2.16295&  $-$10707& 0.055&  36\\
C& 1548&  4896.7& 2.16284&  $-$10717& 0.162&  62& & 4896.7& 2.16285&  $-$10716& 0.082&  53\\
D& 1551&  4865.0& 2.13712& $-$13162& 0.115& 100& & 4865.3& 2.13731& $-$13144& 0.024&  62\\
D& 1548&  4856.5& 2.13689& $-$13184& 0.209& 164& & 4856.6& 2.13694& $-$13179& 0.075& 111\\
E& 1551&  4850.6& 2.12788& $-$14045& 0.180& 147& & 4850.4& 2.12775& $-$14056& 0.114& 142\\
E& 1548&  4842.4& 2.12776& $-$14056& 0.245& 154& & 4842.8& 2.12799& $-$14034& 0.133& 121\\
\\
 & 1551&  4966.3& 2.20246&  $-$6987& 0.024&  27.1& &     ---&     ---&      ---&   ---&   ---\\
 & 1548&  4958.1& 2.20247&  $-$6986& 0.042&  27.2& &     ---&     ---&      ---&   ---&   ---\\
 & 1551&  4852.5& 2.12909& $-$13930& 0.077&  14.2& &     ---&     ---&      ---&   ---&   ---\\
 & 1548&  4844.5& 2.12910& $-$13928& 0.123&  17.3& &     ---&     ---&      ---&   ---&   ---\\
 & 1551&  4829.0& 2.11395& $-$15380& 0.090&  23.6& &     ---&     ---&      ---&   ---&   ---\\
 & 1548&  4821.0& 2.11394& $-$15382& 0.163&  26.1& &     ---&     ---&      ---&   ---&   ---\\

\hline
\end{tabular}
\end{minipage}
\end{table*}

For the variable systems A and E, we record the REWs and FWHMs in 
Table 2 after removing contributions from unrelated 
non-variable lines (see Figs. 2, 4 and 5 below). This was achieved 
by extrapolating the A and E 
line profiles across the tops of the much narrower unrelated 
features. We are guided in this by the fact that the 
unrelated features did not vary between 2006.64 and 2008.67. 
Thus we can accurately decompose the blends into variable 
and non-variable components. 

In the lower resolution SDSS and MMT spectra, the \civ\ systems 
A+B+C and D+E are severely blended together. We therefore measure 
only the total REWs in these blends for comparison to the Keck and 
VLT measurements. We also correct the total SDSS and MMT REWs 
for small contributions from the non-variable lines that are blended 
with systems A and E (as measured from the Keck and VLT spectra). 
Figure 3 plots the total \civ\ REWs in systems A+B+C and D+E 
at each observed epoch. The vertical 
bars in the figure are estimates of the 
1$\sigma$ uncertainties associated with each REW measurement. 
These errors are dominated by the pseudo-continuum 
placement. We estimate the uncertainties by making repeated 
measurements of the REWs using different plausible continuum 
heights. 

One surprising result in Figure 3 is that the REWs differ 
dramatically between the Keck 2006.62 and MMT 2006.71 spectra 
obtained only $\sim$1 month apart in the observed frame. 
These REW differences are also evident from direct 
comparisons of the normalized spectra (not shown). 
They have a formal significance of 3$\sigma$ for systems A+B+C 
and  6$\sigma$ for D+E (based on the uncertainties shown in 
Figure 3). We carefully reviewed our pseudo-continuum 
placements across these lines and find 
no evidence for uncertainties larger than those depicted 
in Figure 3. Nonetheless, we consider this evidence for dramatic 
short-term REW changes to be tentative because of the inherent 
difficulties in comparing the strengths of weak absorption 
lines in spectra with vastly different resolutions. In this case, 
the much shallower appearance of the blended lines in the lower 
resolution data makes them more susceptible to uncertainties 
in the pseudo-continuum placement that might be caused, for example, 
by small changes in the underlying emission line spectrum. 
We also note that significant line variations did not 
occur between our much more precise VLT measurements 
obtained over a similar period of $\sim$1.5 months in 2008 (Table 1).

 \begin{figure}
 \includegraphics[scale=0.47]{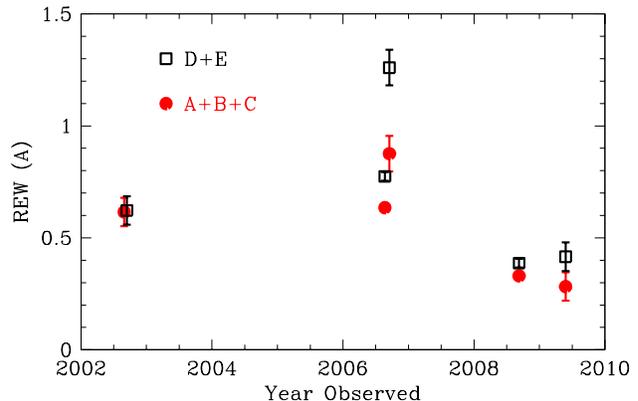}
 \vspace{-15pt}
 \caption{The total rest equivalent widths (REW) of the variable \civ\ 
 systems A+B+C (filled red circles) and D+E (open black squares) 
 are shown for each 
 year observed. The vertical bars indicate approximate 1$\sigma$ 
 uncertainties in the REW measurements. 
   }
\end{figure}

Figures 4 and 5 compare the normalized high-resolution 
Keck and VLT spectra across 
the variable \civ\ and \ovi\ systems. The \ovi\ lines are 
severely blended with unrelated features in the \lya\ forest. 
Nonetheless, the \ovi\ outflow lines are clearly identified 
by their variability and their distinctive  
profiles and redshift agreement with \civ . 
Notice that the \ovi\ lines are both stronger and broader 
than the \civ\ features, with variable absorption that appears 
to extend across a wider range of velocities compared to \civ . 
\begin{figure*}
 \includegraphics[scale=0.78]{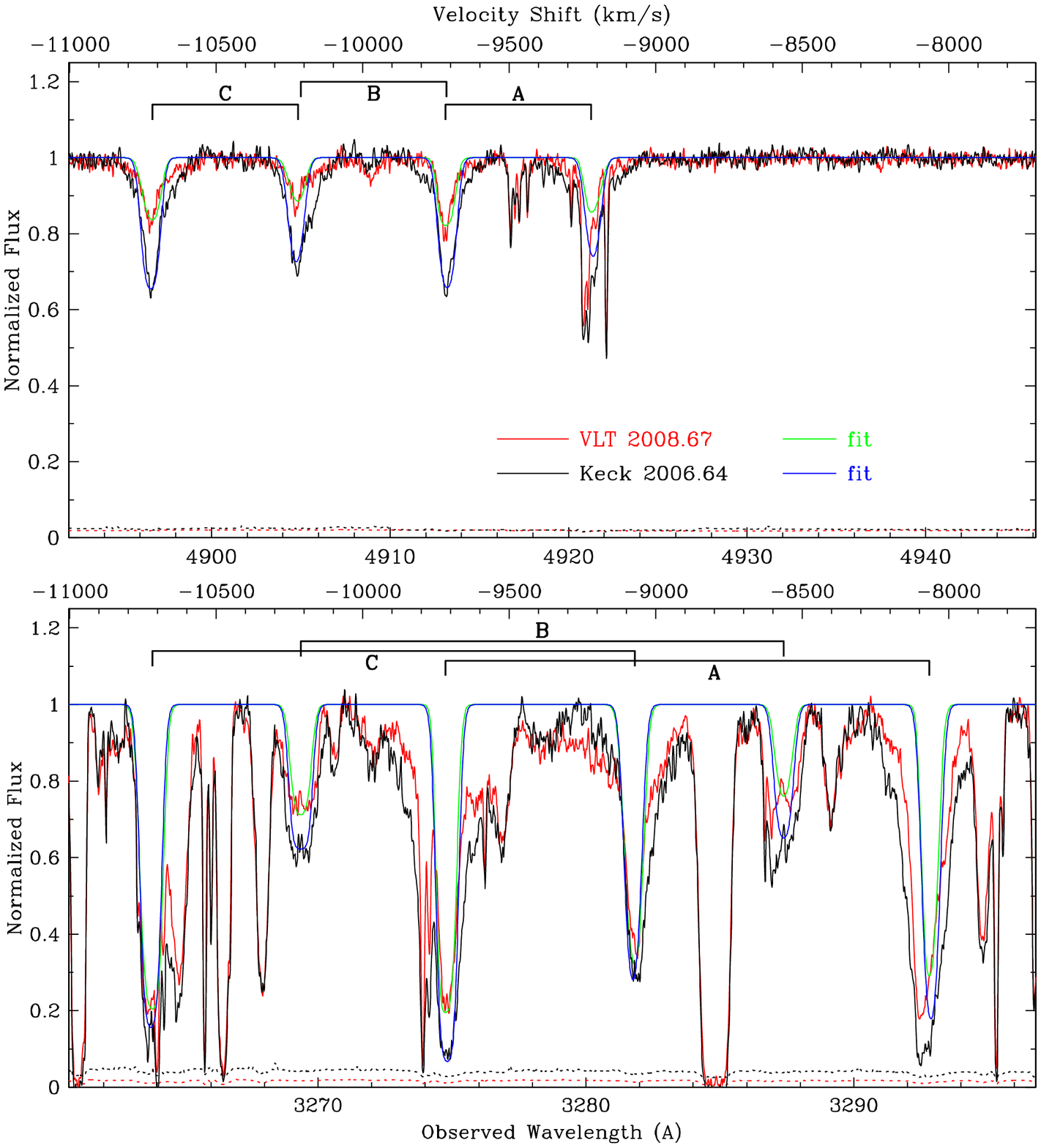}
 \vspace{-5pt}
 \caption{Normalized spectra showing the \civ\ (top panel) and \ovi\ 
 (bottom) doublets in the variable systems A, B and C as observed in 
 the Keck 2006.64 (black curve) and VLT 2008.67 (red) data. The 
 doublet positions are shown by the brackets 
 above the spectra. Many unrelated absorption lines are present 
 in the \ovi\ plot caused by the \lya\ 
 forest and by a system of extremely narrow H$_2$ features in an 
 intervening galaxy at $z=2.059$ \citep{Malec10}. The 
 velocity scale across the top of each panel applies to the short 
 wavelength lines in the doublets, \civ\ 1548 and \ovi\ 1032, 
 relative to the emission redshift, $z_e = 2.278$. The smooth 
 blue and green curves are fits to the variable lines in the Keck 
 and VLT data, respectively, as described in \S3.3. The dotted 
 curves across the bottom are the 1$\sigma$ uncertainties. 
   }
\end{figure*}
\begin{figure*}
 \includegraphics[scale=0.78]{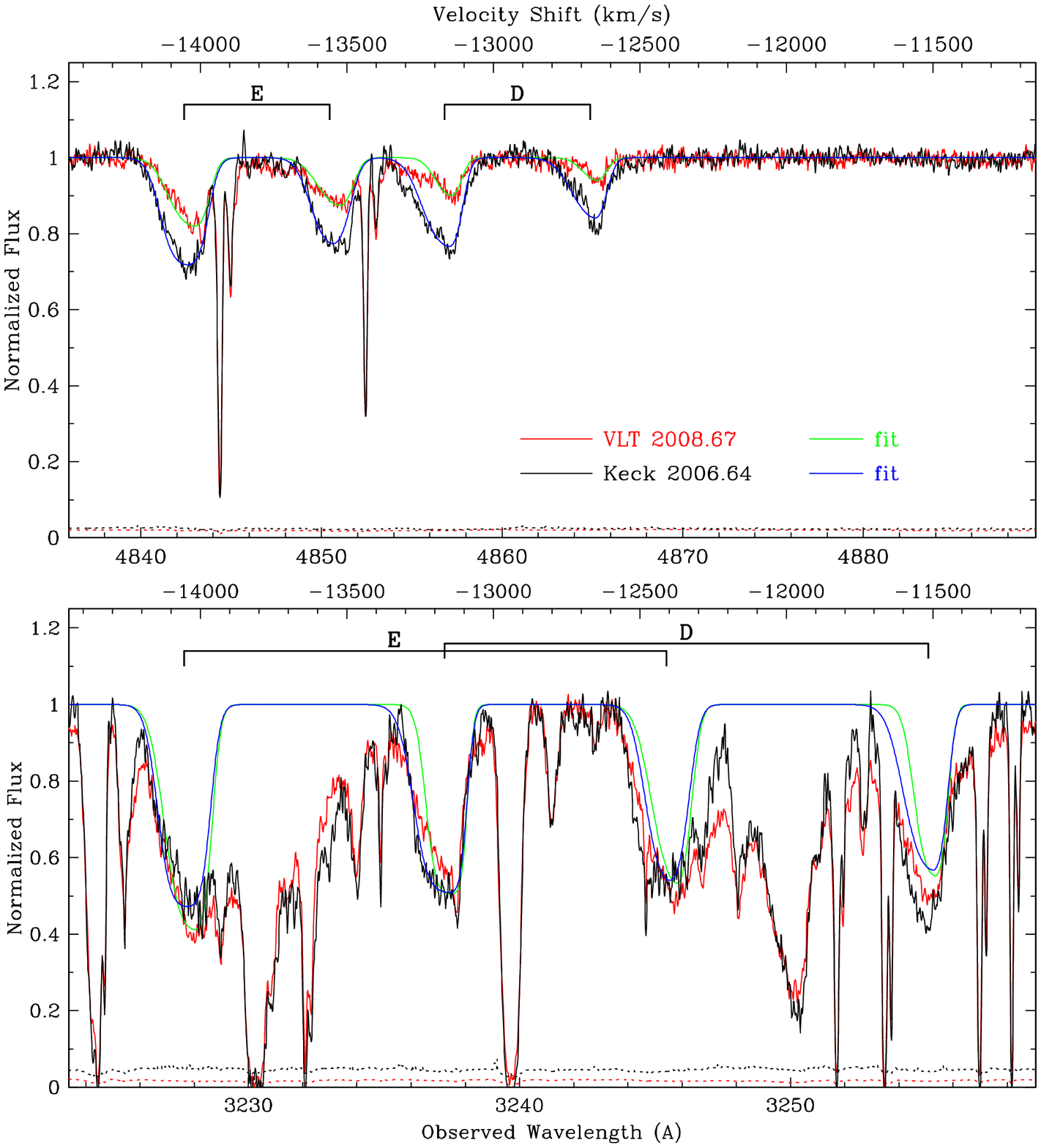}
 \caption{Normalized spectra showing the \civ\ (top panel) and \ovi\ 
 (bottom) doublets in the variable systems D and E. See Figure 4 caption.}
\end{figure*}

Figure 6 shows the Keck and VLT spectra across \lya\ 
in the variable systems A, B and C. \lya\ is clearly detected only in 
system C, based on its good match to the redshift, absorption profile 
and variability characteristics of \civ . 
\lya\ is significantly not present in system B, 
while its strength in the other systems (including D and E, not 
shown) is poorly constrained because of line blending problems.

 \subsection{Covering Factors \& Column Densities}

The smooth green and blue curves in Figures 4--6 are fits to 
the variable line profiles. 
These fits assume that at each velocity, $v$, 
a spatially homogeneous absorbing 
medium covers a fraction, $0<C_{v}\leq 1$, of a spatially 
uniform emission source. In this situation, the line intensities seen 
by a distant observer are given by  
\begin{equation}
{{I_{v}}\over{I_c}} \ = \ (1-C_{v}) + C_{v}\,e^{-\tau_{v}}
\end{equation}
where 
$I_c$ is the unabsorbed source intensity and $\tau_{v}$ and $C_v$ 
are the line optical depth and covering fraction, respectively 
\citep[see][for more general formulations]{Ganguly99,Hamann99,Hamann04}. 
We can solve Equation 1 for $\tau_v$ and $C_v$ at each $v$ 
using the intensity ratios of lines 
within multiplets, such as \civ\ \lam\lam 1548,1551, where the 
optical depth ratio is known from the atomic physics,   
in this case $\tau_{1548}/\tau_{1551} \approx 2.0$ 
\citep[see also][]{Hamann97,Barlow97}. For the resonance 
transitions discussed here, and assuming the ions that are entirely 
in their ground states, the ionic column densities are given by
\begin{equation}
N_{ion} \ = \ {{m_e c}\over{\pi e^2f\lambda_o}}\int\tau_v\,dv
\end{equation}
where $f$ and $\lambda_o$ are the line oscillator strength and 
laboratory wavelength, respectively \citep{Savage91}.   

Previous studies have shown that $\tau_v$ and $C_v$ 
can both have complex velocity-dependent behaviors that 
differ significantly between lines and between ions 
\citep{Hamann97,Barlow97,Ganguly99,Hamann01,Hamann04,Gabel05,
Gabel06,Arav05,Arav08}. 
Unfortunately, blending problems prevent us from obtaining 
$v$-dependent solutions to 
Equations 1 and 2 for most lines in the variable 
systems A--E. We therefore adopt a parameterized approach 
that involves i) fitting the \civ\ profiles with constant 
values of the covering fractions, $C_v = C_o$, across the line 
profiles, and then ii) shifting and scaling the \civ\ fits 
to place constraints on other lines. This procedure also has the 
advantage of avoiding potentially large 
uncertainties in $C_v$, $\tau_v$ and therefore $N_{ion}$ that 
can be introduced by noise, e.g., in the line wings, when 
solving Equation 1 point-by-point at each $v$ across the line 
profiles. The reliability of our fitting procedure is discussed 
further at the end of this section  
\citep[see also][]{Simon10}. 

Our fits assume gaussian $\tau_v$ profiles 
with potentially different values of the 
doppler width parameter on either side of line center. This 
functional form provides flexibility to account for some of the 
asymmetry in the measured lines, while still preserving a simple 
relationship between $N_{ion}$ and the line 
center optical depth, $\tau_o$, namely, 
\begin{equation}
N_{ion} \ = \ 3.34\times 10^{14} \left({{b_b+b_r}\over{f\lambda_o}}\right)
\tau_o \ \ \ \ {\rm cm}^{-2}
\end{equation}
where $\lambda_o$ is the lab wavelength in \AA , and $b_b$ and $b_r$ 
are the blue and red side doppler parameters, respectively, 
in \kms . Each doublet pair is forced to have the same doppler 
parameters, velocity shift and a 2:1 ratio in the line optical 
depths. 

Our initial attempts to fit the \civ\ profiles 
using a $\chi^2$ minimization technique led to  
spurious results. We found that manual adjustments and iteration 
by trial and error provide more flexibility and a much 
better understanding of the uncertainties in situations like this 
where the line profiles are irregular and the blending 
is potentially severe (e.g., in the \civ\ systems A+B+C). 

The fit results in Figures 4 and 5 show that the 
asymmetric gaussians provide a good match 
to the \civ\ profiles in systems D and E, but they are not able 
to produce simultaneously the narrow cores and extended wings in 
systems A, B and C. We choose the ``best'' fits by giving 
highest priority to matching the line cores because they 
are most important for the accuracy of key 
parameters, $\tau_o$, $C_o$ and $N_{ion}$. 
Note that the amount of \civ\ absorption contributed by the 
fully blended system B is highly uncertain. The line ratios 
across the blend A+B+C indicate that the amount of 
\civ\ absorption in system B is less than A and C. 
Repeated attempts to fit \civ\ in the three systems A, B and C  
simultaneously\footnote{There is no straightforward way 
to combine the intensities of blended lines with 
partial covering because the absorbing regions  
can have different values of $C_v$ across 
different spatial locations in front of the emission source.  
Our treatment assumes 
the maximum amount of spatial overlap between absorbing 
regions. For example, if absorbing regions 1 and 2 have covering 
fractions $C_1 > C_2$ and line optical depths $\tau_1$ and $\tau_2$, 
respectively, then the total covering fraction is $C_1$. 
The total line optical depth is $\tau_1 +\tau_2$ in the spatial 
overlap region with covering fraction $C_2$, and $\tau_1$ 
in the non-overlap region with covering fraction $C_1-C_2$. 
See \cite{Hall03,Hall07} for more examples and illustrations.} 
indicate that system B might contribute up to $\sim$30\% of 
the total REW in the lines at this location. 
However, the data are also consistent with no 
\civ\ contribution from system B at all. Therefore none 
is included in the final fits shown in Figure 4 (also 
Table 3 below).  

\begin{figure}
 \includegraphics[scale=0.6]{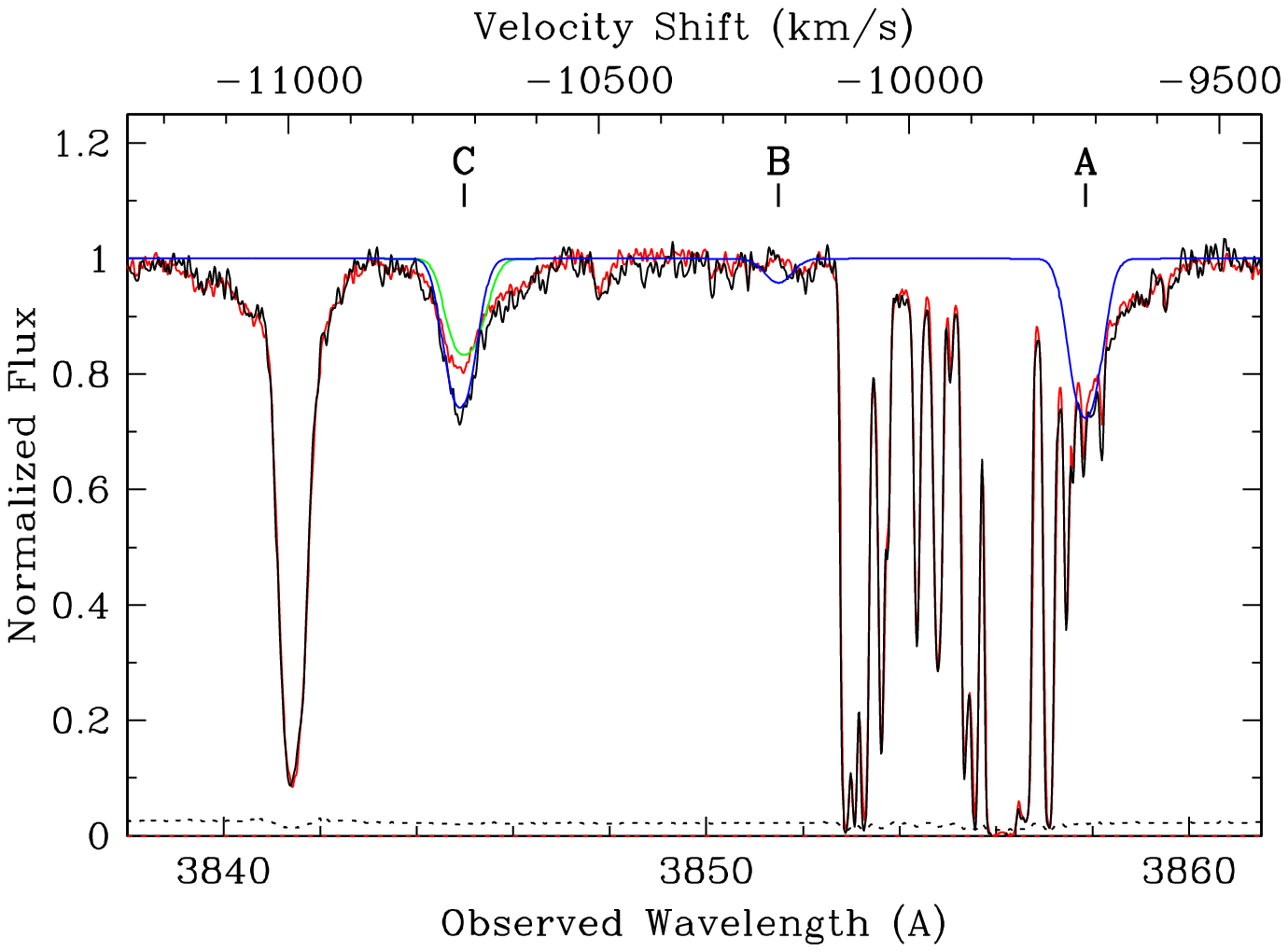}
 \vspace{-8pt}
 \caption{Normalized spectra obtained at Keck (2006.64, black curve) 
 and the VLT (2008.67, red) showing the \lya\ line in 
 the variable systems A, B and C. 
 Other absorption lines, not labeled, are unrelated features formed 
 in intervening clouds or galaxies. See Figure 4 caption.}
\end{figure}

Table 3 lists the \civ\ fit parameters derived from the Keck (2006.64) 
and VLT (2008.67) spectra. Columns 1 and 2 indicate the system 
names and the observation years. Columns 3 and 4 give the 
redshifts, $z_o$, and velocity shifts, $v_o$, of 
the gaussian centers. (Note that these are not centroid positions 
unless the profile is symmetric.) Columns 
5 and 6 list the blue and red side doppler parameters. 
Column 7 gives the line-center optical depths for 
the stronger \lam 1548 transition. Column 8 lists the 
covering fractions, and Column 9 provides the ionic column densities 
from Equation 3. We estimate 1$\sigma$ uncertainties in these 
quantities by trial and error experiments with different fit parameters and 
different pseudo-continuum levels. The uncertainties in $v_o$, $b_r$ and $b_b$ 
should be $\sim$3 \kms\ in systems A and C and roughly double that in the 
broader systems D and E. The covering fractions are in all cases 
well constrained, with 
errors of $<$10\%. The optical depths and column densities have 
nominal 1$\sigma$ uncertainties around 25\%. Including a 
maximum possible contribution from system B would lead to lower 
values of $\tau_o$ and $N_{ion}$ in systems A and C by about 30\%.

\begin{table*}
 \centering
\begin{minipage}{130mm}
  \caption{\civ\ Fit Results}
  \begin{tabular}{@{}ccccccccc@{}}
  \hline

System & Year& $z_o$& $v_o$ & $b_b$& $b_r$& $\tau_o$& $C_o$& $N_{ion}$ \\
 & & & (\kms )& (\kms )& (\kms )& & & (10$^{13}$ cm$^{-2}$)\\
\hline

A& 2006.64&  2.17350 &  $-$9709  & 30   & 30   &  2.3 & 0.38 & 15.7\\
A& 2008.67&  2.17344 &  $-$9715  & 30   & 25   &  2.8 & 0.19 & 17.5\\
C& 2006.64&  2.16278 & $-$10723  & 30   & 30   &  2.7 & 0.37 & 18.4\\
C& 2008.67&  2.16284 & $-$10717  & 30   & 35   &  1.0 & 0.20 & 12.5\\
D& 2006.64&  2.13727 & $-$13148  & 95   & 45   &  1.5 & 0.30 & 23.8\\
D& 2008.67&  2.13733 & $-$13142  & 65   & 40   &  0.8 & 0.18 & 9.54\\
E& 2006.64&  2.12790 & $-$14043  & 80   & 60   &  2.8 & 0.30 & 44.5\\
E& 2008.67&  2.12816 & $-$14018  & 90   & 50   &  1.7 & 0.22 & 27.0\\

\hline
\end{tabular}
\end{minipage}
\end{table*}

Next we transfer the \civ\ fits to other lines in the variable 
systems by keeping the kinematic parameters $v_o$, $b_b$ and $b_r$ 
fixed while scaling only $\tau_o$ and $C_o$, as needed. 
For system B, we adopt $v_o = -10,210$ \kms\ and $b_r=b_b=30$ \kms\ 
based on our efforts to add a system B contribution to the \civ\ blend. 
This simple shift-and-scale approach is essential for dealing 
with non-detections and blends in the \lya\ forest. It is also 
justified by the good results shown for \ovi\ and \lya\ in 
Figures 4-6. However, as noted in \S3.2 above, the \ovi\ absorption 
appears to span a wider range of velocities than \civ . 
Our fits ignore this additional \ovi\ absorption because i) it is 
poorly measured due to blending problems, and ii) we want to limit 
our comparisons between \civ\ and \ovi\ to the same kinematic 
gas components anyway, for the ionization and abundances analysis 
in \S3.6 and \S3.7 below.  

Table 4 lists the fit parameters derived for the other lines. 
The \ovi\ doublets measured in 2006.64 and 2008.67 provide 
useful constraints on both $C_o$ and $N_o$. 
Notice that the \ovi\ covering 
fractions are larger than \civ\ (c.f., Table 3). Also note that 
several of the \ovi\ doublets are saturated, with intensity ratios 
close to unity (see Figs. 4 and 5). For these systems, Table 4 
lists lower limits on the column densities that correspond 
to  $\tau_o = 4$ in the stronger \ovi\ \lam 1032 line. The 
values of $N_{ion}$ listed for \ovi\ system C and system 
E in 2008.67 (only) come from fits indicating $\tau_o <4$. 
However, we could conservatively consider all of the \ovi\ 
column densities to be lower limits because of the 
blending problems in the \lya\ forest. We estimate 
1$\sigma$ uncertainties in the \ovi\ results by repeating 
the fits with different continuum placements while keeping $z_o$ 
and the doppler parameters fixed, as described above. We find that 
the errors in $C_o$ are $<$15\% while the errors in $N_{ion}$ are 
roughly 30\%. We conclude that the differences in the derived 
covering fractions between \ovi\ and \civ , and the smaller 
changes in the \ovi\ covering fractions compared to the changes 
in \civ\ (c.f., Tables 3 and 4), both have a large significance.

\begin{table}
 \centering
\begin{minipage}{70mm}
  \caption{Addtional Fit Results}
  \begin{tabular}{@{}ccccc@{}}
  \hline

  Line& System & Year& $C_o$& $N_{ion}$ \\
 & & & & (10$^{13}$ cm$^{-2}$)\\
\hline
\ovi & A& 2006.64&      0.95 & $>$58.4\\
\ovi & A& 2008.67&      0.82 & $>$53.5\\
\ovi & B& 2006.64&      0.38 & $>$58.4\\
\ovi & B& 2008.67&      0.30 & $>$58.4\\
\ovi & C& 2006.64&      0.87 & ~~51.1\\
\ovi & C& 2008.67&      0.82 & ~~55.4\\
\ovi & D& 2006.64&      0.50 & $>$136\\
\ovi & D& 2008.67&      0.50 & $>$102\\
\ovi & E& 2006.64&      0.54 & $>$136\\
\ovi & E& 2008.67&      0.63 & ~~92.0\\
\\
\siiv & A& 2006.64&    ---& $<$0.30\\
\siiv & B& 2006.64&    ---& $<$0.30\\
\siiv & C& 2006.64&    ---& $<$0.30\\
\siiv & D& 2006.64&    ---& $<$0.60\\
\siiv & E& 2006.64&    ---& $<$0.60\\
\\
\nv & A& 2006.64&    ---&  ---\\
\nv & B& 2006.64&    ---&  $<$8.5\\
\nv & C& 2006.64&    ---&  $<$18\\
\nv & D& 2006.64&    ---&  ---\\
\nv & E& 2006.64&    ---&  $<$72\\
\\
\hi & A& 2006.64&    ---& $<$5.2\\
\hi & B& 2006.64&    ---& $<$0.6\\
\hi & C& 2006.64&    ---&    4.8\\
\hi & C& 2008.67&    ---&    7.7\\ 
\hi & D& 2006.64&    ---& $<$6.4\\
\hi & E& 2006.64&    ---& $<$23\\
\\
\ciii & A& 2006.64&    ---& $<$5.0\\
\ciii & B& 2006.64&    ---& $<$3.0\\
\ciii & C& 2006.64&    ---& $<$3.0\\
\ciii & D& 2006.64&    ---& $<$15\\
\ciii & E& 2006.64&    ---& $<$15\\

\hline
\end{tabular}
\end{minipage}
\end{table}

The remaining lines listed in Table 4 are either non-detections 
(\siiv\ and \ciii ) or severe blends (\nv\ and \hi ) where we cannot 
use line intensity ratios to determine both $\tau_o$ and $C_o$. We 
therefore fix $C_o$ along with the kinematic parameters  
and scale only $\tau_o$ to derive constraints on $N_{ion}$. 
Figure 6 shows examples of these fits applied to 
\lya\ in systems A, B and C. We adopt values of $C_o$ from  
\civ\ (Table 3) in all cases except for \lya\ 
in system B, which has no \civ\ fit result, and \nv , which 
has measured lines/blends too deep to be consistent with $C_o$ 
in \civ . For these latter cases we adopt the covering fractions from 
\ovi . Two of the entries for \nv\ in Table 4 are blank because 
the blending is too severe to yield meaningful results. 
The remaining three \nv\ systems (B, C and E) 
are listed as upper limits in Table 4 because of blending 
uncertainties. We note, however, that there do appear to be 
real \nv\ lines present with redshifts and profiles similar 
to \civ . The upper limits on $N_{ion}$ listed for \hi\ in Table 4 
use the best available constraints from either \lya\ or \lyb . 
The specific values of the \hi\ column density given 
for system C come from the \lya\ fits shown in Figure 6 without 
additional constraints from \lyb . The 
errors in these \hi\ results might be as large as a 
factor of $\sim$2 due mainly to the uncertainties in $C_o$. 

As a further check of our fitting analysis, 
we derive velocity-dependent values of 
$\tau_v$ and $C_v$ for the two best-measured cases: the \civ\ 
lines in systems D and E observed in 2006.64. 
Specifically, we calculate the average line intensities in 
velocity bins 25 \kms\ wide and then solve 
Equation 1 for $\tau_v$ and $C_v$ in each bin. 
Figure 7 compares the derived values of $1-C_v$ 
to the observed line profiles. The vertical bars running through 
the filled circles in Figure 7 represent 1$\sigma$ 
uncertainties in $C_v$ caused by photon statistics \citep{Hall03}. 
However, in the blue wings of both systems D and E there are 
several velocity bins where the doublet ratios indicate 
$\tau_v\ll 1$ and the standard analysis yields $C_v=1$. 
The actual values of $\tau_v$ and $C_v$ are poorly constrained 
in these cases because the right side of Equation 1
reduces to $1-C_v\tau_v$ in the $\tau_v\ll 1$ limit and, therefore,   
the line intensities constrain only the product 
$C_v\tau_v$ instead of $C_v$ and $\tau_v$ 
separately. For example, in the velocity bin 
centered at $v= -13,250$ \kms\ in \civ\ system D, the 
data are consistent with solutions ranging from $1-C_v =0$ and 
$\tau_v(1548)=0.14$ to $1-C_v\approx 0.5$ and $\tau_v(1548)
= 0.28$. Figure 7 shows both the maximum and minimum ($1-C_v=0$) 
results, connected by a vertical bar, for each of the optically 
thin data points. The data at these velocities are compatible 
with any value of $1-C_v$ within the ranges shown. 

\begin{figure}
\begin{center}
 \includegraphics[scale=0.65]{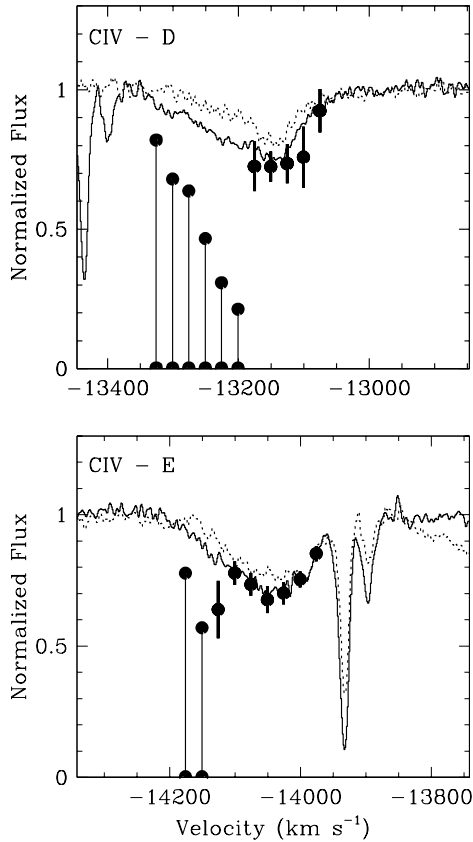}
 \vspace{-8pt}
 \caption{\civ\ line profiles measured in 2006.64 in systems D (top 
 panel) and E (bottom), compared to derived values of $1-C_v$ 
 (filled circles), on a velocity scale relative to 
 $z_e = 2.278$. The solid and dashed curves show the 
 stronger (\lam 1548) and weaker (\lam 1551) doublet members, 
 respectively, after smoothing the spectra 3 times with a binomial 
 function. The vertical bars on the filled circles are 1$\sigma$ 
 uncertainties. For the velocity bins with $\tau_v\ll 1$ (in the 
 blue line wings), two filled circles at each velocity represent the 
 full range of $1-C_v$ values consistent with the data.
   }
\end{center}
\end{figure}

The values of $C_v$ and $\tau_v$ derived from the point-by-point 
analysis near line center agree well with 
$C_o$ and $\tau_o$ obtained above from the gaussian fits 
(Table 3). The point-by-point results in Figure 7 
show further that $C_v$ changes significantly  
across the line profiles. In particular, 
the shape of the intensity profile across the core of system E  
appears to be governed largely by the behavior of $C_v$ with 
$\tau_v>2$. In system D, there appears to be an abrupt 
change in $C_v$ around $-$13,200 \kms , between a line core that 
has $\tau_v > 1$ and $C_v\approx 0.28$ and a blue wing with 
$\tau_v\ll 1$ and $C_v\approx 1$. Integrating $\tau_v$ 
over all of the velocity bins shown in Figure 7 (as per 
Equation 2 but excluding the lowest $v$ point in the red wing of both 
systems because they formally yield $\tau_v\gg 1$ due to noise) 
indicates $N_{ion} =  2.3\pm 0.3\times 10^{14}$ \cmN\ for system D 
and $3.8\pm0.3 \times 10^{14}$ \cmN\ for system E. These column 
densities are in excellent agreement with the gaussian fit 
results in Table 3. 

We conclude with a comment about the assumption of 
spatial uniformity in the absorbing regions. Some studies of 
partial covering in quasar absorption lines indicate that the 
the actual absorbing structures 
are inhomogeneous, with a range of 
$\tau_v$ values at each $v$ across the 
projected area of the emission source  
\citep[][and refs. therein]{Hamann01,deKool02,Hamann04,Arav08}. 
The evidence for this is significantly different covering 
fractions in different lines/ions. While these data can provide 
important constraints on the nature of the absorber inhomogeneities, 
there is not enough information to determine uniquely the 
2-dimensional spatial distributions of $\tau_v$ in each line. 
The situation is further complicated by the fact that the 
background light sources, e.g., the accretion disk and/or the BELR, 
have their own spatial non-uniformities that can include a 
wavelength-dependent size. A good strategy, therefore, is to consider
 the observational consequences of different plausible $\tau_v$ 
spatial distributions \citep{deKool02,Hamann04,Arav05,Arav08}. 
In one study, extensive simulations using different functional 
forms for the $\tau_v$ spatial distribution show that Equation 1 
(which assumes a top hat $\tau_v$ distribution) yields good 
approximations to the spatially-averaged values of $\tau_v$ 
and $N_{ion}$ in more general inhomogeneous absorbing media 
\citep{Hamann04}. 
Therefore, the results derived from Equation 1 are appropriate 
for making spatially-averaged estimates of the ionization and 
metal abundances (\S3.6 below). 
They are also useful for characterizing the covering fractions even for 
complex absorbers. In particular, the values of $C_v$ derived 
from Equation 1 correspond roughly to the amount of coverage 
by absorbing material with $\tau_v \ga 1$ in an inhomogeneous 
medium \citep{Hamann04}. 

\subsection{Absorbing Region Sizes}

The partial covering discussed in \S3.3 implies 
that the absorbing structures are not 
much larger than the background light 
source. (The absorbers could be smaller than the emission 
source, but partial covering by much larger clouds would require 
unphysically sharp edges and extremely unlikely spatial 
alignments.) Figure 2 shows that the variable \civ\ systems 
lie in the far blue wing of the \civ\ BEL, where 
the flux is dominated by continuum emission. 
In principle, the line absorption could occur in front of 
both the UV continuum source and the BELR, 
with potentially different covering fractions for each region 
\citep{Ganguly99}. However, the depths of the lines below 
the quasar continuum (Figure 2) imply that the \civ\ partial covering 
results pertain mostly or exclusively to the 
continuum source. This conclusion is confirmed by the 
\ovi\ lines, which have even deeper absorption troughs 
and no significant underlying BELR flux. 

The extent to which the absorbers cover the BELR are thus undetermined 
by our data. This ambiguity introduces a small additional uncertainty 
in $C_v$ derived for the \civ\ lines because the BELR contributes roughly 
10\% of the total flux at those wavelengths (Figure 2). 
Our analysis in \S3.3 assumed implicitly that $C_v$ is 
the same for all emission sources beneath the absorption 
lines. However, if the BELR is not covered at all, the continuum 
covering fractions would be roughly 10\% larger than the values 
listed for \civ\ in Table 3. Conversely, if the \civ\ absorber 
completely covers the BELR, the continuum covering fractions would 
be $\sim$20\% smaller than the results listed in Table 3 
\citep[see equations in][]{Ganguly99}. This latter situation (complete 
coverage of the BELR with partial covering of the continuum source) 
seems highly unlikely given that the \civ\ BELR is of order 20 times 
larger than the continuum source at 1550 \AA\ (\S2.1). 
Hereafter, we will assume that all of the derived covering fractions 
apply only to the continuum source. 

Comparing the covering fractions in \civ\ and \ovi\ 
(Tables 3 and 4) to the theoretical diameters of the continuum 
source at 1550 \AA\ and 1034 \AA , respectively (\S2.1), 
indicates that the \civ\ and \ovi\ absorbing regions both have 
characteristic sizes of order $\sim$0.01-0.02 pc (as measured 
in 2006.64). It is important to keep 
in mind, however, that this characteristic size is actually 
just an upper limit on the true sizes of absorbing structures 
in the flow. In particular, partial covering might be 
caused by patchy distributions of many smaller sub-structures 
\citep[see \S4.4 below, also][]{Hamann04}. 

\subsection{Outflow Dynamics}

The five distinct outflow systems span a range of velocities 
from $-$9710 to $-$14,050 \kms\ and \civ\ line widths from 
FWHM $\approx$ 62 to 164 \kms\ (Table 2). The \ovi\ lines, 
representing higher ionization material,  
appear somewhat broader than \civ\ in terms of both their 
FWHMs and broad weak wings (see Figs. 4 and 5). 
The outflow systems have a natural division into 
two groups, A+B+C and D+E, based on similarities in the line 
strengths, profiles and velocity shifts (see also Fig. 1). 
One remarkable characteristic of these systems overall is their 
small velocity dispersions compared to the high flow speeds, 
e.g., FWHM/$v\sim 100$. This in marked contrast to BAL 
outflows, which typically have FWHM/$v\sim 1$ 
\citep{Weymann91,Korista93}. 

The kinematic stability of the 
narrowest systems A, B and C between 2006.64 and 
2008.67 places an upper limit on the radial 
acceleration of $\la$3 \kms\ yr$^{-1}$ in the quasar 
frame. We obtain this estimate from the measurements 
in Table 3 and by careful comparisons of simultaneous fits 
to all three narrow systems (in \civ\ and \ovi ) in both epochs. 
(The centroids of the broader systems D and E did 
change slightly between observations, but this appears to be the 
result of profile changes rather than a true velocity shift; 
Figure 5.) For comparison, the gravitational acceleration due 
to $M_{BH}$ at the radius of the \civ\ BELR (\S2.1) is roughly 
200 \kms\ yr$^{-1}$. Therefore, the outward 
driving force is either in a delicate 
balance with gravity or, more likely, the absorbing 
gas is located in a low-gravity environment 
farther from the black hole than the \civ\  
BELR. A radial distance of $R \ga 5$ pc is needed for the local 
gravity to be less than the observed acceleration limit. At this 
distance, the observed outflow speeds vastly exceed the local 
gravitational escape speed. Thus it appears that the absorbing 
gas is gravitationally unbound and coasting freely 
at speeds near the outflow terminal velocity. 

There is an apparent 
occurrence of ``line-locking" between the narrow \civ\ systems 
A, B and C. In particular, the velocity differences between systems 
A-B and B-C are a close match to the \civ\ doublet separation of   
$\Delta v = 498$ \kms . (Although the 
system B lines are entirely blended in 
\civ , their velocity is confirmed by our fits to the \ovi\ 
lines shown in Figure 4.) 

Line-locking is usually interpreted 
as evidence for radiative acceleration 
\citep{Foltz87,Braun89,Srianand00,Srianand02,Ganguly03}. 
Cases of multiple line-locks between systems with narrow 
profiles, as in J2123-0050, provide the best evidence for 
physically locked systems because they are difficult to 
explain by chance alignments of lines in the spectrum. 
However, it is not known if radiative forces can 
actually lock doublets like \civ\ together in real quasar 
outflows. The most plausible scenario was outlined by 
\cite{Braun89}: Consider two blobs of gas 
moving along the same radial path in a radiatively 
accelerated outflow. The faster blob is farther from the 
emission source and experiencing a {\it greater} outward 
acceleration. When 
the velocity difference between the blobs matches the \civ\ 
doublet separation, $\Delta v\approx 498$ \kms , 
the \lam 1551 line in the outer blob falls in 
the shadow cast by \lam 1548 in the slower inner blob. This leads to 
diminished acceleration in the outer blob and a velocity lock 
between the two blobs at the doublet separation. Line-locking   
might even contribute to the formation of blobs because the 
localized dip in the radiative force downstream from the 
inner blob creates a natural collecting area for outflow gas 
\citep{Ganguly03}. This scenario does not require that 
the \civ\ lines dominate the overall acceleration, only that  
the acceleration transferred to the outer blob via the \lam 1551 
line is at least as large as the acceleration 
{\it difference} between the two blobs in the absence of 
shadowing (i.e., outside of the $\Delta v\approx 498$ \kms\ 
line overlap situation). The appearance of a double line-lock 
in \civ , as in J2123-0050, might also be facilitated by the 
\nv\ absorption 
doublet whose separation, $\Delta v = 964$ \kms , is close to 
the velocity difference between systems A and C. In any case, 
the feasibility of this line-lock interpretation 
has not been demonstrated in any quasar outflow. 

Here we note simply that if radiative forces did indeed lock the \civ\ 
doublets in J2123-0050, then the absorbing clouds must 
be moving almost directly toward us. This is because the 
velocity difference between line-locked blobs should match  
the \civ\ doublet separation along the flow 
direction. If the flow is not aimed directly at us, we should 
observe velocity differences between the systems that are 
smaller than the doublet separation. 
Our measurements of the line centroids (Table 2) and our fits 
to the profiles, including \ovi\ in system B (Table 3 and Figure 4), 
indicate that the velocity differences between systems 
A, B and C are within $\sim$20 \kms\ of $\Delta v = 498$ \kms . 
This implies that the absorber trajectories are within 
$\sim$16$^o$ of the purely radial (line-of-sight) direction. 

\subsection{Physical Conditions}

The strong detections of \ovi\ compared to \civ\ absorption, 
and the non-detections of lower ionization lines like \siiv\ and 
\ciii\ (Tables 3 and 4), indicate that the degree of ionization 
is high and approximately the same in all of the variable 
systems. The Appendix below describes numerical simulations 
using the code Cloudy \citep{Ferland98} of 
the ionization and physical conditions in clouds that are in 
photoionization equilibrium with a quasar continuum source. 
Comparing those results (Figure A1) to the most stringent limit 
on the observed $N(\ciii )/N(\civ ) \la 6$ column density 
ratio (system C in 2006.64) indicates that the dimensionless 
ionization parameter is at least $\log U \ga -0.8$. 
The measured ratios of $N(\ovi )/N(\civ ) \ga 3$ provide 
a very similar constraint on $\log U$ if we make 
the additional assumption that the O/C 
abundance is approximately solar. 

This lower limit on the ionization combined with  
the average measured \hi\ column density in 
system C, $N(\hi )\sim 6\times 10^{13}$ \cmN\ (Table 4),  
implies that the minimum total hydrogen column density 
in that system is $N_H \ga 3\times 10^{18}$ \cmN\ 
(see Figure A1). In \S4.3 below, we argue that the line 
changes were caused by changes in the degree 
of ionization and the most likely value of the 
ionization parameter is $\log U\sim -0.4$,  
near the peak in the \ovi\ ion fraction. Using that result, 
our best estimate of the actual total 
column density in system C is 
of order $N_H \sim 10^{19}$ \cmN . Given the similarities 
between the systems A-E, we crudely estimate that the 
total column density in all five outflow absorbers is 
$N_H \sim 5\times 10^{19}$ 
\cmN .

The more conservative requirement for $\log U\ga -0.8$ in a 
photoionized gas at $R \ga 5$ pc (\S3.5) from the quasar places 
an upper limit on the volume density of $n_H\la 2\times 10^8$ \cmn\ 
(see Equation A1). A lower limit on the density can be derived 
by noting that, in photoionization equilibrium, 
changes in the ionization state require at least 
a recombination time given by 
$t_r \approx  (\alpha_r n_e)^{-1}$, 
where $n_e$ is the electron density and $\alpha_r$ is the 
recombination rate coefficient. Given $\alpha_r \approx 
10^{-11}$ cm$^3$ s$^{-1}$ for either 
\ovii\ $\rightarrow$ \ovi\ or \ovi\ $\rightarrow$ \ov\ 
at a nominal gas temperature of $T_e \approx 20,000$ K 
\citep[see][]{Arnaud85,Hamann95}, the observed \ovi\ 
variability in $\leq$0.63 yr indicates  
a minimum density of $n_e\sim n_H\ga  5000$ \cmn . This 
density combined with the ionization constraint $\log U\ga -0.8$  
yields a maximum distance of $R\la 1.1$ kpc (Eqn. A1). 

The density and distance constraints based on the 
recombination time ($n_H\ga  5000$ \cmn\ and $R\la 1.1$ kpc) 
might not apply if the ionization changes occurred out of 
equilibrium. In particular, ionization {\it increases} caused by 
a rise in the ionizing flux could occur much faster than the 
recombination time. The most likely explanation for the line 
changes between 2006.64 and 2008.67 is, indeed, an 
ionization increase (\S4.3 below). However, there is weak 
evidence for both weakening and strengthening of the \civ\ lines 
in J2123-0050 (Fig. 3), which is typical variable outflow lines 
in quasars \citep{Hamann95,Misawa07c,Narayanan04,Hamann97b,
Paola10,Gibson10,Capellupo10}. If this up and down behavior of 
the line strengths represents changes in the ionization   
\citep[see also][]{Misawa07c}, then recombination must be involved. 
Continued high-resolution monitoring is needed to confirm that 
conclusion specifically for J2123-0050. However, even if the line 
variations in this object occurred because of clouds crossing our 
view of the continuum source, basic assumptions about 
the crossing speeds would require a radial distance much smaller 
than the 1.1 kpc limit derived from the recombination time (\S4.3). 
Therefore, we adopt $R\la 1.1$ kpc as a firm upper limit. 

\subsection{Metal Abundances}

We estimate the C/H metal abundance from the average 
measured ratio of $N(\civ )/N(\hi )\sim 2.5$ 
in system C (Tables 3 and 4) and an ionization correction 
based on $\log U\sim -0.4$ (see Eqn. A2 and Fig. A1). This 
indicates [C/H] $\sim 0.3\pm 0.2$, where the square brackets 
have their usual meaning of log abundance relative to solar 
and the 1$\sigma$ 
uncertainty includes an estimate of the errors in both 
the column densities and the ionization correction. 
The uncertainties related to the ionization are 
small because the most likely ionization parameter, 
$\log U\sim -0.4$, is near the peak 
in the \civ\ ion fraction curve where the  
correction factor, $f(\hi )/f(\civ )$, is not 
sensitive to the specific value of $\log U$ across a 
fairly wide range. A careful inspection of the numerical 
results depicted in Figure A1 shows that 
$f(\hi )/f(\civ )$ stays constant within $\pm$0.1 dex 
across a range in $\log U$ from $-$1.4 to 0.0.  
Moreover, this range in $\log U$ encompasses the 
minimum value of $f(\hi )/f(\civ )$ at $\log U = -0.75$ 
(Table A1). If we apply the minimum correction factor 
$\log (f(\hi )/f(\civ )) = -3.80$ (Table A1) to the system 
C column densities, we derive a lower limit on the 
metal abundance, [C/H] $\ga 0.2$, which is close to our 
best estimate above but completely independent of the 
ionization uncertainties 
\citep[see also][]{Hamann97d}.

\section{Discussion}

Section 3 presented measurements and analysis of five distinct 
outflow NAL systems in J2123-0050 having \civ\ line widths 
of FWHM $\sim$ 62 to 164 \kms\ and velocity shifts from 
$v\sim -9710$ to $-$14,050 \kms\ in the quasar rest frame. 
These systems appear to be physically 
related based on their roughly similar line strengths, kinematics, 
ionizations, line-of-sight covering factors, and coordinated 
variabilities. They provide a wealth of information 
about a particular NAL outflow that is valuable 
for comparison to other work on BAL and mini-BAL outflows. 
Here we discuss some additional 
results and implications. Please see \S5 below for an overall 
summary.

\subsection{Location Summary}

The formation of the variable NALs in a quasar outflow 
is confirmed by all three properties of 
line variability, partial covering and resolved 
profiles that are smooth and broad compared to thermal 
line widths. Moreover, if we accept that the lines form 
somewhere in the quasar environment based on these properties, 
then the only plausible location is a quasar-driven 
outflow because the velocity shifts are much too 
large to be explained by other near-quasar environments 
such as a galactic starburst-driven 
wind or a nearby galaxy in the same cluster as the 
quasar \citep{Prochaska09,Rupke05}. By matching the 
variability to a recombination time, we place an 
upper limit on the distance between the absorber and the quasar 
of $R\la 1.1$ kpc (\S3.6). At the opposite extreme, 
constraints on the acceleration suggest  
that the NAL gas is coasting freely at a distance $R\ga 5$ 
pc (\S3.5). Thus the full range of plausible distances 
is $5\la R\la 1100$ pc. However, if the 
absorbing structures are small blobs or filaments created 
in the inner flow, their most likely location is near 
the $\sim$5 pc minimum radius because such structures will  
travel just a few pc before dissipating (if there is no 
external confinement. \S4.4 below). 

Finally, our estimate of the metallicity, [C/H] $\sim 0.3\pm 0.2$ 
with a lower limit of [C/H] $\ga 0.2$, is also 
consistent with the formation of these NALs in a quasar 
outflow. Super-solar metallicities are extremely rare in 
intervening absorption line systems, having been observed 
so far only in a few Lyman-limit and super-Lyman-limit 
systems \citep{Prochaska06,Prochter10}, but they are 
typical of near-quasar environments 
based on estimates from the broad emission lines 
\citep{Hamann99,Dietrich03,Nagao06} and from other 
well-studied cases of NAL outflows 
\citep[][Simon, Hamann \& Pettini 2010, in prep.]{D'Odorico04,Gabel05,Gabel06,Arav07}.

\subsection{The Outflow Origins of Quasar NALs}

It is interesting to note that the outflow NALs in 
J2123-0050 are unresolved and indistinguishable from 
cosmologically intervening lines in medium resolution 
spectra like the SDSS. They are much 
too narrow to be identified with an outflow using one of 
the absorption line indices designed for this purpose 
\citep{Weymann91,Trump06}. Their velocity shifts are also well 
above the nominal $\vert v\vert < 5000$ \kms\ cutoff used to 
define ``associated'' absorption lines (AALs), which are likely 
to have a physical relationship to the quasar based on their 
statistical excess at $z_a\approx z_e$ compared to $z_a\ll z_e$ 
\citep{Weymann79,Foltz86,Nestor08,Wild08}. Without high 
resolution spectra or multi-epoch 
observations to test for variability, the outflow NALs in 
J2123-0050 would be mistakenly attributed to cosmologically 
intervening gas. 

The outflow systems are, in fact, surrounded in 
the J2123-0050 spectrum by other narrow \civ\ lines that 
did not vary, do not have partial covering and almost 
certainly do form in unrelated intervening material (Figure 1). 
One of these intervening systems is at a smaller 
velocity shift than the outflow lines (at $v\sim 6987$ \kms ), 
while another is blended directly with the 
outflow system E (see also Table 2 and Figure 5). 
This mixture of lines in the same spectrum clearly 
demonstrates that velocity shift alone is a poor 
indicator of the outflow versus intervening origin 
of narrow absorption lines \citep[see also][]{Simon10}. 

The secure outflow origin of the NALs in J2123-0050 lends 
anecdotal support to recent claims that a significant fraction of 
narrow \civ\ systems, even at large velocity shifts, form in 
quasar outflows. For example, \cite{Richards01a} argued that 
$\sim$36\% of high-velocity NALs originate in outflows based on a 
correlation between the numbers of these systems detected and the 
radio properties of the background quasars. \cite{Misawa07a} 
estimated that 10-17\% of narrow high-velocity \civ\ systems 
belong to quasar outflows based on the incidence of partial 
covering measured in high-resolution spectra. Similar results were 
obtained more recently by Simon et al. (2010, in prep.). 
\cite{Nestor08} showed that the statistical excess of \civ\ NALs 
near the quasar redshift extends out to at least $\sim$12,000 \kms , 
e.g., well beyond the nominal cutoff of 5000 \kms\ for ``associated'' 
absorption lines.  
They argue that $\ga$43\% of NALs stronger than REW = 0.3 \AA\ 
in the velocity range $750\la v\la 12,000$ \kms\ originate in 
quasar outflows \citep[also][]{Wild08}. In their analysis of mostly 
weaker systems, Simon et al. (2010, in prep.) find that roughly 
20-25\% of \civ\ NALs in this velocity range exhibit partial covering. 

The variable lines in J2123-0050 are unique in that they are (so 
far) the highest velocity NALs known to originate in a quasar  
outflow based on all three indicators of line 
variability, partial covering and super-thermal line widths. 
Their closest analogues in the literature are a complex of NALs 
at somewhat lower speeds, $v\sim 9500$ \kms , in the redshift 
2.5 quasar HS~1603+3820 \citep{Misawa05,Misawa07c}, and an isolated broader  
(FWHM $\sim 400$ \kms ) outflow system at $v\sim 24,000$ \kms\ in 
the $z_e\approx 2.5$ quasar Q2343+125 \citep{Hamann97e}. 

\subsection{What Caused the NAL Variability?}

The key properties of the NAL variability in J2123-0050 
are 1) the line variations were well coordinated between 
the five systems, each with larger changes in \civ\ than \ovi , 
2) changes in the NAL strengths were accompanied by 
roughly commensurate changes in the line-of-sight 
covering fractions, and 3) the variability time is $\leq$0.63 yr 
in the quasar rest frame. 

Changes in the covering fractions would seem to have a natural 
explanation in clouds moving across our lines of sight to 
the background emission source. This explanation 
is, in fact, favored by some recent studies of BAL variability 
\citep{Gibson08,Gibson10,Hamann08,Capellupo10}. 
However, in J2123-0050, line changes induced by crossing clouds 
appear unlikely because the coordinated variations in five distinct 
NAL systems would require highly coordinated movements between 
five distinct absorbing structures in the outflow 
\citep[also][]{Misawa05}. 
Moreover, crossing clouds would need to traverse a significant 
fraction of the emission source diameter in $\leq$0.63 yr to 
cause the observed line variations. In \S3.4 we 
argued that the relevant emission source 
is the accretion disk 
with diameter $D_{1550}\sim 0.026$ pc at 1550 \AA . 
Crossing $\geq$15\% of that length in $\leq$0.63 yr, to explain 
the observed changes of order 0.15 in the \civ\ covering fractions,  
would require transverse speeds $v_{tr}\ga 6,000$ \kms . 
This seems unrealistic given that the radial flow speeds are 
not much larger, e.g., $v\sim -9710$ to 14,050 \kms . 
The maximum transverse speeds we might 
expect in a quasar outflow launched from a rotating accretion disk 
are of order the virial speed at the 
absorber location. A virial speed consistent with 
$v_{tr}\ga 6000$ \kms\ in J2123-0050 would require that the absorber 
resides at a radial distance $\la$2.4 pc from the SMBH. This would 
place the absorbers inside of the minimum radius of 5 pc  
estimated above from the lack of radial acceleration (\S3.5). 
It would also contradict the 
evidence from line-locking that the flow we observe is 
traveling primarily in a radial direction (\S3.5). 

A more likely explanation for the coordinated line variations is 
global changes in the ionization caused by fluctuations 
in the quasar's continuum flux. This situation could produce the 
observed changes in the covering fractions (\S3.3) if the 
absorbers are spatially inhomogeneous, i.e., with a distribution of 
column densities across the emission source, such that the  
projected areas with $\tau_v\ga 1$ in a given line change with 
the ionization state of the gas \citep{Hamann04}. 
The observed decrease in the \civ\ line strengths between 
2006.64 and 2008.67, accompanied by smaller or negligible 
changes in \ovi , are consistent with an overall  
increase in the ionization parameter if the starting point was 
near the peak in the \ovi\ ion fraction. In particular, 
an ionization parameter initially near $\log U\sim -0.4$ in 
2006.64 and then increasing by a few tenths of a dex by 2008.67 
would cause a decline in the \civ\ line strengths and 
column densities consistent with the observations (Table 3), 
while the \ovi\ remains nearly constant (see Figure A1). 

This scenario requires that the quasar's far-UV 
continuum flux changed by a factor of roughly two in 
$\leq$0.63 yr (rest). Little is known about the far-UV 
variability properties of luminous quasars. 
Nonetheless, this type of continuum change seems plausible 
given that 1) near-UV flux variations as large as 
30-40\% have been observed in other luminous quasars 
on times scales of months 
\citep[rest,][]{Kaspi07}, and 2) flux changes in the 
far-UV could be much larger than the near-UV because shorter 
continuum wavelengths generally exhibit larger 
amplitude fluctuations 
\citep{Krolik91,VandenBerk04}. 

The recent SDSS stripe 82 
survey shows specifically that the near-UV continuum variability 
of J2123-0050 is typical of luminous quasars \citep{Schmidt10}. 
This survey includes roughly 60 epochs of imaging photometry 
collected over a $\sim$7 year period that overlaps with our 
spectroscopic observations. Examination of these data 
for J2123-0050 (kindly provided by Kasper Schmidt) shows 
continuum variations up to $\sim$0.27 
magnitudes (or $\sim$28\%) on observed time scales of 
$\sim$2 years in the $r$ band ($\sim$2000 \AA\ rest). 
Moreover, the $r$-band flux increased by this amount between 
roughly 2005.8 and the last stripe 82 data point in 2008.0, which 
is consistent with an our conclusion that the ionization increased 
between the Keck and VLT observations in 2006.64 and 2008.67. 

The main caveat to interpreting the NAL variabilities  
in terms far-UV flux changes is that there {\it might} 
have been a large NAL variation between 2006.64 (Keck) 
and 2006.71 (MMT), i.e., in just $\sim$8 days in the quasar 
frame (Figure 3). 
If this extreme short-term variability can be confirmed, 
it might be difficult to explain via factor of $\sim$2 changes 
in the far-UV emission from the accretion disk 
\citep[but see also][]{Misawa07c}. 

\subsection{Outflow Structure}

The outflow NALs in J2123-005 require at least five 
distinct absorbing structures that have similar physical 
conditions (ionization, column densities, covering fractions, 
\S3.3 and \S3.4) and roughly similar kinematics (lines dispersions 
and velocity shifts, \S3.5). The partial covering evident in 
all five systems indicates that the absorbers 
are each characteristically smaller than the quasar's UV 
continuum source, e.g., $\sim$0.01-0.02 pc across. 
All five absorbers must also be spatially inhomogeneous to 
explain the observed changes in the covering fractions 
if those changes are, indeed, caused by changes in the quasar's 
ionizing flux (\S4.3). 

The absorbing structures might be discrete blobs 
or filaments created by instabilities in the inner 
flow, perhaps resembling the clumpy structures seen in some 
numerical simulations of BAL winds \citep{Proga04}.
If this assessment is correct, then their most likely location 
is near the minimum radius of $R\sim 5$ pc inferred from 
the lack of acceleration (\S3.5). This is because small 
blobs or filaments have short survival times against 
dissipation if there is no external pressure confinement. The 
characteristic absorber diameter of $d\sim 0.015$ pc (\S3.4) combined 
with the doppler parameters $b\sim 70$ \kms\ of the broader 
systems D and E (Table 3) indicate dissipation times of 
order $t_{dis}\sim d/b \approx 210$ yr. This is much longer than 
our  observation timeline (Table 1) and therefore not a factor 
in the line variability. However, gas blobs moving  
radially at $v\sim 13,500$ \kms\ (for systems D and E) will 
travel a distance of just $\sim$3 pc before dissipating. 
Therefore, they should be located within $\sim$3 pc of their 
point of origin. If that was the inner flow near the accretion disk 
\citep[e.g., at radii not much greater than a 
few times $R_{BELR}\sim 0.65$ pc, \S2.1,][]{Murray97,Proga00}, 
then the absorbers should reside near the $R\sim 5$ pc 
minimum radius (\S4.1). 

One puzzling constraint on the internal structure of these 
absorbers comes from the high densities. At $R\sim 5$ pc, 
the densities need to be near the upper end of the 
estimated range, $10^8\ga n_H\ga 5000$ \cmn , to avoid 
over-ionization by the intense quasar radiation field (\S3.6). 
However, this entire density range is much larger than 
the average gas density, $\left<n_H\right>\sim 215$ \cmn , 
expected for absorbing clouds with total column density 
$N_H\sim 10^{19}$ \cmN\ (\S3.6) and characteristic size  
$\sim$0.015 pc. This disparity implies that each 
of the five absorbers is composed of dense sheets or 
sub-structures that have radial thickness much smaller 
than their characteristic {\it transverse} size of $\sim$0.015 pc. 
The combined radial thickness of these sheets or 
sub-structures in each absorbing region is just 
$N_H/n_H\sim 10^{11}$ cm if they are at $R\sim 5$ pc, 
or $\sim$$2\times 10^{15}$ cm at $R\sim 1100$ pc.

Some early studies of BALs presented similar arguments for the 
existence of dense sub-structures with an overall 
small volume filling factor in quasar outflows \citep{Weymann85,Turnshek95}. 
Those studies also recognized that the creation and survival of 
such small sub-structures present a serious theoretical 
challenge for outflow models \citep{deKool97}. 
These concerns about sub-structure 
effectively ended when \cite{Murray95} and \cite{Murray97} 
proposed an attractive alternative, namely, that the flows 
could be spatially smooth and continuous if there is a 
shielding medium at the base of the flow that 
blocks most of the quasar's intense X-ray and far-UV 
radiation. With this shielding medium in place, the outflow gas behind 
the shield should be able to maintain the observed moderate degrees of 
ionization at much lower gas densities, thus allowing the flow 
to have an overall smooth spatial distribution. Subsequent X-ray 
observations supported this scenario by showing that BAL quasars 
are heavily absorbed in X-rays 
\citep{Green96,Mathur00,Gallagher02,Gallagher06}. 

However, J2123-0050, like other NAL and mini-BAL outflow quasars, 
does not have strong X-ray absorption (\S1 and \S2.2) and 
therefore it does not have a significant X-ray/far-UV 
shield \citep[see also the calculations in][]{Hamann10}. 
Small dense sub-structures do appear 
necessary to explain the moderate degree of ionization at the 
inferred flow locations. Thus the debate about the existence of 
small sub-structures in quasar outflows remains open. 
These sub-structures might resemble the small-scale clumps identified 
recently in one detailed observational study of partial 
covering in a BAL outflow \citep{Hall07}. 

\subsection{Acceleration \& Radiative Shielding: Comparisons 
to BAL Outflows}

The most remarkable aspect of the NAL outflow in
J2123-0050 is the degree to which it contrasts with BAL 
flows. For example, it has high speeds and ionizations very 
similar to BALs, but the NALs we measure are about a hundred 
times narrower and a hundred times weaker (in REW) than 
typical BALs \citep[c.f., Table 2 and][]{Weymann91,Korista93}. 
Our estimate of the total column density, 
$N_H\sim 5\times 10^{19}$ \cmN , is also 1-2 orders of 
magnitude less than the best available estimates obtained from  
BALs \citep{Hamann98,Arav01,Hamann02}. The weak or 
absent X-ray absorption in J2123-0050 \citep{Just07} 
is typical of other NAL and mini-BAL outflow sources 
\citep{Gibson10,Chartas09,Misawa08}, but it too is very 
different from the strong X-ray absorption found in 
BAL quasars \citep{Mathur00,Gallagher02,Gallagher06}. 
 
This disparity in the X-ray properties between 
BAL, mini-BAL and NAL outflows has important implications 
for outflow physics. As noted in \S4.4, the prevailing 
models of radiatively-driven BAL outflows have a shielding medium 
at the base of the flow that can absorb the quasar's 
ionizing far-UV/X-ray flux and thus prevent the over-ionization 
of the outflow gas behind the shield \citep{Murray95,Murray97}. 
Moderate ionizations in the outflow gas are essential to 
maintain significant opacities and facilitate the radiative 
acceleration to high speeds. However, the absence 
of strong X-ray absorption in J2123-0050 shows that high 
speeds and moderate ionizations can occur in quasar 
outflows {\it without} significant radiative shielding. 

One way to reconcile these results is by a 
unified scheme where the outflows measured 
via BALs, NALs and mini-BALs coexist at different 
spatial locations. In situations like this 
\citep[e.g.,][]{Ganguly01,Chartas09}, the acceleration 
of the NAL and mini-BAL gas might occur in the shielded 
BAL environment (near the accretion disk plane) before 
it moves to its observed location (farther above the disk) 
where the shielding is negligible. The discrete appearance of 
NALs and mini-BALs in quasra spectra might be due to their 
formation in discrete blobs or filaments that are created via 
instabilities along the ragged edge of the main 
BAL outflow \citep{Ganguly01,Hamann08}. However, it is not 
clear there can be enough vertical force to push these 
blobs significantly away from the disk plane and out of 
their natural radial trajectories 
away from the UV continuum source. A more basic problem 
with this picture is that 
NALs and mini-BALs have degrees of ionization similar to BALs 
even though they are not behind the radiative shield that is 
alleged to be important for controlling the outflow ionization. 
If the NAL and mini-BAL gas is accelerated in a shielded BAL 
environment where it has moderate BAL-like ionizations, how does 
it maintain those moderate BAL-like ionizations after it emerges 
from the shielded BAL zone? 

The obvious answer from our analysis of J2123-0050 
(\S3.6 and \S4.4) is that the outflow ionization 
is regulated mostly by locally high gas densities and not by a 
radiative shield. High densities could keep the 
ionization low enough for radiative acceleration 
even if there is no shielding at all. Perhaps this is the case 
generally in quasar outflows, including the BALs. However, 
high densities also lead to the uncomfortable conclusion 
that the outflows are composed of small sub-structures 
that have an overall small volume filling factor (\S4.4). 
The physical processes that might create or maintain 
these sub-structures are not understood 
\citep{deKool97}. Nonetheless, the evidence for their 
existence in J2123-0050 is at least consistent 
with high densities playing a more important role in the 
acceleration physics than radiative shielding.  

\subsection{Energetics \& Feedback}

Here we estimate the total mass and kinetic 
energy in the J2123-0050 outflow to determine if it can 
plausibly be important for feedback to the host galaxy's 
evolution (\S1). If the flow geometry is approximately like 
part of a thin spherical shell, then its total mass is given by   
\begin{equation}
M \ \approx \ 14\; \left({{Q}\over{10\%}}\right)\left({{N_H}\over{5\times 
10^{19}\,{\rm cm}^{-2}}}\right)\left({{R}\over{5\,{\rm pc}}}\right)^2 
~~~ {\rm M}_{\odot}
\end{equation}
where $N_H\approx 5\times 10^{19}$ \cmN\ is our best guess at the 
total column density in all five absorbing systems (\S3.6), 
$R\approx 5$ pc is a likely radial distance (\S4.1 and \S4.4),  
and $Q$ is the global covering fraction of the outflow, 
i.e., the fraction of 4$\pi$ steradians covered by the flow 
as seen from the central continuum source \citep{Hamann00iop}. 
The value of $Q$ is not constrained 
by our data, but $Q\sim 10\%$ is a reasonable guess 
based on estimates of the detection rates of high-velocity NAL 
outflows (\S4.2). The outflow kinetic energy defined by $K=Mv^2/2$ 
is therefore  
\begin{equation}
K \ \approx \ 2\times 10^{52} \; \left({{M}\over{14\,{\rm M}_{\odot}}}\right)
\left({{v}\over{12,000\, {\rm km/s}}}\right)^2~~~{\rm ergs}
\end{equation}
where $v\approx 12,000$ \kms\ is roughly representative of all 
five systems. 

To estimate the mass loss rate and kinetic energy 
luminosity, we would need to make additional assumptions about 
the duration and radial structure of the flow 
\citep{Hamann00iop}. However, if we simply divide $K$ from Equation 
5 by a characteristic flow time, $t_{flow}\sim R/v\sim 410$ yr 
(for $R\sim 5$ pc and $v\sim 12,000$ \kms ), 
we obtain a time-averaged kinetic luminosity of   
$\left<L_K\right>\sim 2\times 10^{42}$ ergs s$^{-1}$ and a ratio 
relative the quasar's bolometric photon luminosity of 
only $\left<L_K\right>/L \sim 2\times 10^{-6}$ (\S2.1). This 
estimate is much less than the ratio of $L_K/L\sim 5$\% 
expected if the outflow is to be important for feedback 
\citep{Scannapieco04,DiMatteo05,Prochaska09}. 
We also note that the total kinetic energy given by Equation 5 
is equivalent to the non-neutrino output of only  
$\sim$20 type II supernovae (at $K\sim 10^{51}$ ergs each). 

The total mass and kinetic energy could be much larger if the 
outflow gas is near the maximum radius, $R\sim 1100$ 
pc, derived in \S3.6. In that case, Equations 4 and 5 indicate 
$M\sim 7\times 10^5$ \msun\ and $K\sim 10^{57}$ ergs, 
respectively. However, this value of $K$ is still far less than 
the binding energy of gas in a massive galaxy 
($\sim$$10^{59}$-$10^{60}$ ergs), and the increased flow time 
at $R\sim 1100$ pc leads to a time-averaged kinetic energy yield, 
$\left<L_K\right>/L \sim 4\times 10^{-4}$, that is still 
two orders of magnitude below 
the fiducial mark of $L_K/L\sim 5$\% needed for feedback. 
Therefore, at any of the plausible absorber locations, the 
kinetic energy yield from this NAL outflow is much too small to 
be important for feedback to the host galaxy's evolution. 

\section{Summary}

We discuss multi-epoch spectroscopic observations of five 
outflow NAL systems in the luminous quasar J2123-0050. 
The lines have velocity shifts from $-$9710 
to $-$14,050 \kms\ and \civ\ line widths in the range 
FWHM $\sim$ 62 to 164 \kms\ (as measured in 2006.64, Table 2). 
These are the highest velocity NALs reported so far to have their 
formation in a quasar outflow confirmed by all three indicators 
of line variability, partial covering of the quasar continuum 
source, and smooth super-thermal line profiles. 
The five distinct absorption line systems 
require five outflow structures with similar physical 
conditions, similar sizes (covering fractions) 
and roughly similar kinematics. All five systems have 
stronger absorption in \ovi\ than \civ , while lower ions 
such as \siiv , \ciii\ and \cii\ are significantly absent. 
The observed NAL variabilities were well coordinated between 
the five systems, with changes in the line strengths accompanied 
by nearly commensurate changes in the line-of-sight 
covering fractions. The time scale for significant NAL changes  
is $\leq$0.63 yr in the quasar rest frame. Our analysis 
of these data provides the following additional constraints 
on the outflow properties: 

1) {\it Absorbing Region Sizes.} 
Partial covering of the UV continuum source 
indicates that the \civ\ and \ovi\ absorbing regions 
have characteristic sizes of order $\sim$0.01-0.02 pc 
(\S3.4). 

2) {\it Ionization.} The absence of \ciii\ lines and stronger 
absorption in \ovi\ compared to \civ\ 
in all five systems indicates an ionization 
parameter of $\log U\ga -0.8$ (\S3.6). The most likely 
value is near $\log U \sim -0.4$ based on the different 
variabilities observed in \civ\ and \ovi\ (\S4.3). 

3) {\it Variability.} 
The coordinated nature of the NAL variations occurring in 
$\leq$0.63 yr is best explained by global ionization changes 
in the outflow caused by changes in the quasar's ionizing flux 
(\S4.3). The magnitude of the ionization changes corresponds to 
roughly 0.2-0.3 dex in $U$.

4) {\it Metallicity.} 
The secure detection of \lya\ in one system indicates 
a metal abundance of [C/H] $\sim 0.3\pm 
0.2$, with a lower limit independent of ionization 
uncertainties of [C/H] $\ga 0.2$ (\S3.7). 
These results are consistent with other 
estimates of quasar metallicities 
and they support the claim that the  
variable NALs in J2123-0050 form in a quasar outflow. 

5) {\it Total Column Density.} 
We estimate the total hydrogen column density in system C to be 
$N_H \sim 10^{19}$ \cmN\ based on $\log U \sim -0.4$ 
and the strength of \lya\ 
in that system. From this and the many similarities between the 
variable systems, we infer that the total column density in all 
five systems is very roughly $N_H \sim 5\times 10^{19}$ \cmN\ 
(\S3.6). 

6) {\it Dynamics.} 
An upper limit on the radial acceleration, $\la$3 \kms\ 
yr$^{-1}$, and an apparent double line-lock in the \civ\ systems 
suggest that i) the outflow was radiatively accelerated, 
ii) it is now gravitationally unbound and coasting freely beyond the 
main acceleration zone, and iii) its trajectory is within $\sim$16$^o$ 
of the purely radial (line-of-sight) direction (\S3.5). 

7) {\it Location.} 
The acceleration upper limit suggests that the outflow lines form 
in a low gravity environment $\ga$5 
pc from the central quasar (\S3.5). The minimum gas density needed 
for recombination in $\leq$0.65 yr sets an upper limit 
on the distance of $\sim$1100 pc (in photoionization equilibrium, 
\S3.6). Thus the full range of 
plausible distances is $5\la R\la 1100$ pc. 
If the absorbing structures are blobs or filaments created 
in the inner flow, their short survival times against dissipation 
(in the absence of a confining pressure) indicates that they 
are near the minimum radius of $\sim$5 pc (\S4.4).

8) {\it Density \& Sub-Structure.} 
The gas densities corresponding to distances 
$5\la R\la 1100$ pc are $10^8\ga n_H\ga 5000$ \cmn\ for a gas 
in photoionization equilibrium at $\log U \ga -0.8$ (\S3.6). 
These densities are much larger than the average value 
$\left<n_H\right>\sim 215$ \cmn\ expected from the total 
column density and characteristic size. Therefore,  
each of the five absorbers is composed of 
dense thin sheets or sub-structures with an overall small 
volume filling factor (\S4.4).

9) {\it Radiative Shielding \& Acceleration.} 
The absence of strong X-ray absorption in J2123-0050 
\citep{Just07} is typical of NAL and mini-BAL outflows but 
contrasts markedly with BAL quasars. It 
implies that radiative shielding in the X-ray/far-UV is not 
needed to keep the outflow ionizations moderate (and BAL-like)
and therefore, perhaps, it is not needed to facilitate the 
radiative acceleration of the flow to high speeds. 
We argue that the ionization is controlled, 
instead, by locally high gas densities in small outflow 
sub-structures (\S4.5). 

10) {\it Energetics \& Feedback.} At the most likely radial 
distance near $R\sim 5$ pc, this outflow has a total mass 
of only $\sim$14 \msun\ and kinetic energy $\sim$$2\times 
10^{52}$ ergs. At any of the plausible larger distances noted 
above,  the kinetic energy yield from this outflow 
is still at least two orders of magnitude too small to be 
important for feedback to the host galaxy's evolution (\S4.6). 

\section*{Acknowledgments}

We are grateful to Dan Capellupo, Anton Koekemoer and Leah Simon 
for helpful discussions, and to 
Kasper Schmidt for providing the SDSS stripe 82 
survey data for our analysis in \S4.3. We thank an anonymous 
referee for helpful comments on the manuscript. This work was 
based in part on observations at the W. M. Keck Observatory, which 
is operated as a partnership between the University of California, 
the California Institute of Technology and the National Aeronautics 
and Space Administration, and at ESO VLT at the Paranal 
Observatories under programme ID 081.A-0242. 
FH acknowledges support 
from the Chandra Award TM9-0005X. NK acknowledges support from the 
Department of Science and Technology through a Ramanujan Fellowship.
JXP is supported in part by an NSF CAREER grant (AST-0548180) 
and by NSF grant AST-0908910. MTM thanks the Australian
Research Council for a QEII Research Fellowship (DP0877998).
WU acknowledges financial support from the Netherlands Foundation
for Fundamental Research of Matter (FOM).

\bibliographystyle{mn2e}

\appendix

\section{Photoionization Calculations}

We use the numerical code Cloudy, version 08.00  
\citep[last described by][]{Ferland98}, to compute ionization  
fractions for clouds that are in photoionization equilibrium 
with a quasar 
continuum source. The results have wide applicability to 
quasar outflows and a variety of other near-quasar environments. 
The calculations assume the gas clouds are dust free, at a 
constant density throughout, optically thin in the Lyman 
continuum and geometrically thin such that their radial thickness, 
$\Delta R$, is small compared to their distance, $R$, 
from the ionizing emission source, i.e., $\Delta R/R\ll 1$. 
The optically thin 
assumption ensures that the model clouds have a uniform 
ionization structure throughout. This is appropriate for  
the majority of quasar outflow systems, including J2123-0050, that 
are highly ionized with $N(\hi ) \la 10^{17.2}$ \cmN . 
\citep[Moderate continuum opacities can occur at the \heii\ 
ionization edge for $10^{16.5}\la N(\hi ) \la 10^{17.2}$ \cmN , 
but this has no significant affect on the ionization fractions 
discussed here,][]{Hamann97d}
The calculations assume that the metallicity 
is solar and the total hydrogen density is $n_H = 10^5$ \cmn , 
although the specific values of these parameters do not affect 
the ionization results \citep{Hamann97d}.

The quasar spectrum is a simple segmented powerlaw 
consistent with observational constraints 
across the most important far-UV energies \citep[][and refs. 
therein]{Elvis94,Telfer02,Steffen06,Richards06,Hopkins07}. 
It has the form $f_{\nu}\propto \nu^{\alpha}$ with 
slope $\alpha = -1.0$ from 0.5 eV to 13.6 eV, $\alpha = -2.56$ 
from 13.6 eV to 136 eV, and $\alpha = -1.0$ 
from 136 eV to 136 keV. Sharper cutoffs are applied to the 
high and low energy extremes ($\alpha = 3$ 
below 0.5 eV and $\alpha = -3$ above 136 keV) that have 
no affect on the computed ionizations.  
This spectrum yields $\alpha_{ox} = -1.6$ \citep[appropriate 
for luminous radio-quiet quasars,][]{Strateva05,Steffen06}, and a 
bolometric correction factor at 1450 \AA\ of  
$L\approx 4.4\; \nu L_{\nu}(1450\, {\rm \AA})$ \citep[see also][]{Warner04}. 

We describe the ionization state of the gas in terms of the 
dimensionless ionization parameter, 
$U \equiv \Phi_H/(4\pi c\, n_H R^2)$, where $\Phi_H$ is the 
hydrogen-ionizing photon luminosity of the central source 
\citep{Ferland98}. 
For the continuum shape described above, this yields the 
following expression for the radial distance 
\begin{equation}
R \ = \ 225\left({{\nu L_\nu (1450{\rm \AA })}
\over{10^{47}\,{\rm ergs~s}^{-1}}}\right)^{1/2}
\left({{10^5\,{\rm cm}^{-3}}\over{n_H}}\right)^{1/2}
\left({{0.1}\over{U}}\right)^{1/2}~{\rm pc}~
\end{equation}

Figure A1 shows the calculated ionization fractions for 
\hi\ and several metal ions commonly observed in near-UV 
spectra of quasars, namely, ions of C, N, O and Si. One 
important application of these results to absorption line work 
is derivations of metal-to-hydrogen (M/H) abundance ratios based on 
this relation 
\begin{equation}
\left[{{\rm M}\over{\rm H}}\right] \ = \ 
\log\left({{N({\rm M}_i)}\over{N(\hi )}}\right) + 
\log\left({{f(\hi )}\over{f({\rm M}_i)}}\right) + 
\log\left({{\rm H}\over{\rm M}}\right)_{\odot}
\end{equation}
where (H/M)$_{\odot}$ is the solar abundance ratio by number, 
and $N$ and $f$ are the column densities and ionization fractions, 
respectively, for \hi\ and the metal M in ion stage $i$. 
Ideally, one would apply Equation 2 after comparing the measured 
column densities in several metal ions to the calculated results 
(Figure A1) in order to define $U$ and an appropriate 
ionization correction factor, $f(\hi )/f({\rm M}_i)$. 
However, if ionization constraints from 
the data are not available or not reliable, one can also use 
minimum values of the ionization corrections in Equation A2 to derive 
firm lower limits on the metal-to-hydrogen abundance ratios 
\citep{Hamann97d,Hamann99}. 

Table A1 lists the minimum correction factors for various metal 
ions, $\log (f(\hi )/f({\rm M}_i))$, together with the values of the 
ionization parameter, $\log U$, where these minima occur. This 
information is provided in the table according to the element 
(column 1) and ionization stage II-VI (columns 2-6) for ions 
that are sometimes detected in quasar outflow spectra. Note 
that the minimum correction factors occur near, 
but at larger $U$ than, the peaks in the $f({\rm M}_i)$ curves 
shown in Figure A1. For convenience, we note that 
the most recent solar abundance ratios are $\log($H/M$)_{\odot} = 
3.57$ for carbon, 4.17 for nitrogen, 3.31 for oxygen, 5.55 for 
aluminum, 4.49 for silicon, 6.59 for phosphorus and 4.88 for 
sulfur \citep{Asplund09}. 

\begin{figure}
 \includegraphics[scale=0.37,angle=-90.0]{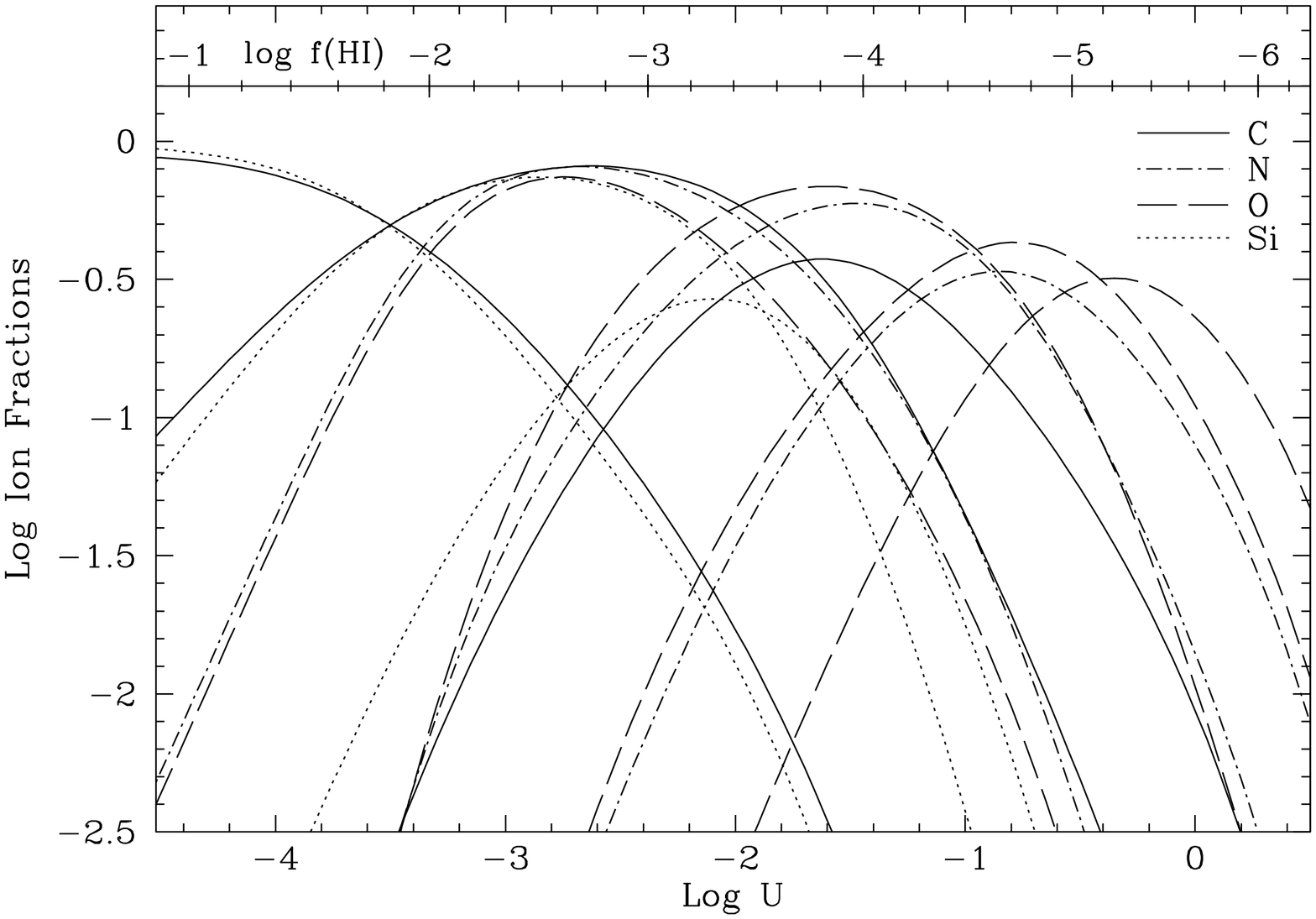}
 \caption{Ion fractions, $f({\rm M}_i)$, for selected 
 stages of the elements C, N, O and Si plotted 
 against the ionization parameter, $\log U$, in optically thin 
 photoionized clouds (see text). From left to right 
 the three curves plotted for carbon are for \cii , \ciii\ and 
 \civ . The remaining curves (left to right) 
 are for nitrogen: \niii , \niv\ and 
 \nv , oxygen: \oiii , \oiv , \ov\ and \ovi , and silicon: 
 \siii , \siiii\ and \siiv . The \hi\ fraction, $\log f(\hi )$, 
 at each $\log U$ is marked along the horizontal bar near the top. 
 }
\end{figure}

\begin{table}
 \centering
 \begin{minipage}{73mm}
  \caption{Minimum Ionization Corrections}
  \begin{tabular}{@{}lccccc@{}}
  \hline
   Element & II & III & IV 
     & V & VI \\
\hline

 ~~~~~~~~&  
 \multicolumn{5}{l}{minimum $\log (f(\hi )/f({\rm M}_i))$:} \\
 Carbon    &   $-$1.75&  $-$3.32&  $-$3.80&  $-$5.39&  ---\\
 Nitrogen  &   $-$1.75&  $-$3.28&  $-$4.15&  $-$4.55&  $-$5.89\\
 Oxygen    &   $-$1.81&  $-$3.02&  $-$4.17&  $-$4.69&  $-$5.06\\
 Aluminum &   $-$2.12&  $-$2.20&  $-$3.36&  $-$4.11&  $-$4.96\\
 Silicon   &   $-$1.65&  $-$2.93&  $-$3.03&  $-$3.89&  $-$4.51\\
 Phosphorus&   $-$1.90&  $-$3.10&  $-$3.34&  $-$3.57&  $-$4.42\\
 Sulfur   &   $-$1.50&  $-$2.95&  $-$3.73&  $-$3.73&  $-$3.97\\
  \\
 ~~~~~~~~& 
 \multicolumn{5}{l}{$\log U$ at minimum $\log (f(\hi )/f({\rm M}_i))$:} \\
 Carbon    &   $-$2.60&  $-$1.55&  $-$0.75&  ~\, 0.30&   ---\\
 Nitrogen  &   $-$3.20&  $-$1.50&  $-$0.75&  $-$0.15&   ~\, 0.65\\
 Oxygen    &   $-$3.15&  $-$1.65&  $-$0.75&  $-$0.15&   ~\, 0.30\\
 Aluminum &   $-$2.60&  $-$2.10&  $-$1.50&  $-$0.70&  $-$0.10\\
 Silicon   &   $-$2.80&  $-$1.95&  $-$1.55&  $-$1.00&  $-$0.35\\
 Phosphorus&   $-$2.60&  $-$1.80&  $-$1.35&  $-$1.00&  $-$0.50\\
 Sulfur   &   $-$2.90&  $-$1.65&  $-$1.20&  $-$0.85&  $-$0.60\\
\hline
\end{tabular}
\end{minipage}
\end{table}

\bsp

\label{lastpage}

\end{document}